\documentclass[pdflatex,sn-nature]{sn-jnl}

\usepackage{graphicx}%
\usepackage{multirow}%
\usepackage{amsmath,amssymb,amsfonts}%
\usepackage{amsthm}%
\usepackage{upgreek}
\usepackage{mathrsfs}%
\usepackage[title]{appendix}%
\usepackage{xcolor}%
\usepackage{textcomp}%
\usepackage{manyfoot}%
\usepackage{booktabs}%
\usepackage{algorithm}%
\usepackage{algorithmicx}%
\usepackage{algpseudocode}%
\usepackage{listings}%
\usepackage{color}%

\theoremstyle{thmstyleone}%

\theoremstyle{thmstyletwo}%

\theoremstyle{thmstylethree}%

\begin{document}

\title[Article Title]{Non-epitaxial perovskite polariton laser diode operating under direct current}

\author*[1]{\fnm{Anatoly P.} \sur{Pushkarev}}\email{anatoly.pushkarev@gmail.com}

\author[1,2]{\fnm{Daria} \sur{Khmelevskaia}}
\equalcont{These authors contributed equally to this work.}

\author[1]{\fnm{Ivan A.} \sur{Matchenya}}
\equalcont{These authors contributed equally to this work.}

\author[1]{\fnm{Stepan A.} \sur{Baryshev}}
\equalcont{These authors contributed equally to this work.}

\author[1]{\fnm{Denis A.} \sur{Sannikov}}

\author[2]{\fnm{Alexey A.} \sur{Ekgardt}}

\author[3]{\fnm{Eduard I.} \sur{Moiseev}}

\author[3]{\fnm{Natalia V.} \sur{Kryzhanovskaya}}

\author[3]{\fnm{Alexey E.} \sur{Zhukov}}

\author[4]{\fnm{Dmitry V.} \sur{Krasnikov}}

\author[5,2]{\fnm{Alexandr A.} \sur{Marunchenko}}

\author[2]{\fnm{Alexey V.} \sur{Yulin}}

\author[4]{\fnm{Albert G.} \sur{Nasibulin}}

\author*[1]{\fnm{Pavlos G.} \sur{Lagoudakis}}\email{p\_lagoudakis@mail.ntua.gr}

\affil[1]{\orgdiv{Hybrid Photonics Laboratory},  
\orgaddress{\street{Bolshoy Boulevard 30, bldg. 1}, \city{Moscow}, \postcode{121205}, \country{Russia}}}

\affil[2]{\orgdiv{School of Physics and Engineering}, \orgname{ITMO University}, \orgaddress{\street{Kronverksky Pr. 49, bldg. A}, \city{St. Petersburg}, \postcode{197101}, \country{Russia}}}

\affil[3]{\orgdiv{International Laboratory of Quantum Optoelectronics}, \orgname{HSE University}, \orgaddress{\street{Soyuza Pechatnikov str. 16}, \city{St.
Petersburg}, \postcode{190008}, \country{Russia}}}

\affil[4]{\orgdiv{Laboratory of Nanomaterials},  
\orgaddress{\street{Bolshoy Boulevard 30, bldg. 1}, \city{Moscow}, \postcode{121205}, \country{Russia}}}

\affil[5]{\orgdiv{Division of Chemical Physics and NanoLund}, \orgname{Lund University}, \orgaddress{\street{PO Box 124}, \city{Lund}, \postcode{22100}, \country{Sweden}}}

\abstract{Reaching lasing in electrically pumped microdevices based on solution-processed semiconductors poses a major scientific and technological challenge. Halide perovskites offer a promising platform for electrical injection~\cite{shi2025ten}, since their optically excited single-crystal cavities~\cite{zhu2015lead,liao2015perovskite,eaton2016lasing} and predesigned~\cite{tian2023perovskite,kkedziora2024predesigned} or postprocessed microstructures~\cite{sun2020lead,huang2020ultrafast} have exhibited low lasing threshold. Indirect electrical pumping of a dual-cavity perovskite laser was recently obtained~\cite{zou2025electrically}, utilizing a well-established technological concept of embedding a high-luminosity light-emitting diode (LED) with a high-gain medium into an integrated device~\cite{yang2008hybrid}. Direct charge-carrier injection into a perovskite LED excited by auxiliary short-, optical-pulses resulted into amplified spontaneous emission~\cite{elkhouly2024electrically}. Other efforts for rational engineering of architectures~\cite{kim2018hybrid, kim2020optically,zhao2021nanosecond,elkhouly2022intense} that allow for high charge-carrier density are still to demonstrate lasing. Here, we develop a novel strategy for achieving direct electrical pumping of a perovskite laser. We integrate a solution-grown CsPbBr$_3$ microplate with chemically inert single-walled carbon nanotube electrodes and embed them into an optical microcavity. By cooling the microdevice down to 8 K at a constant current, a perovskite p-i-n diode is formed that facilitates a balanced carrier injection at high current densities. The perovskite microcavity diode operates in the strong coupling regime, exhibiting polariton lasing under a direct current of 65 $\upmu$A.
}


\maketitle

\section{Introduction}\label{sec1}
Solution-processed semiconductors offer a promising, inexpensive technology to supersede their epitaxial counterparts on the market~\cite{garcia2017solution, saki2021solution, deng2022solution}. Among such optoelectronic devices like light-emitting diodes (LEDs), solar cells (SCs), and photodetectors (PDs), the most challenging for engineering are laser diodes. The latter operate at high current densities that have proven difficult to achieve in organic semiconductors due to their low carrier mobility~\cite{wang2021pursuing}. Furthermore, substantial losses in electrically pumped organic semiconductors stem from the formation of triplet excitons, polarons, and singlet-triplet annihilation among others~\cite{wang2021pursuing}. Although some of these issues were partially addressed for carbazole derivative-based distributed feedback (DFB) microlasers \cite{sandanayaka2019indication}, the examined devices did not operate under continuous current but only for a small number of electrical pulses.

Alternatively, robust solution-processable inorganic materials have surpassed some of the aforementioned limitations~\cite{wang2021pursuing}. An edged-emitting diode with a CdSe/Cd$_{1-x}$Zn$_x$Se/ZnSe$_{0.5}$S$_{0.5}$/ZnS colloidal quantum dots layer operating under high current density up to 1.93 kA cm$^{-2}$ exhibited stable amplified spontaneous emission (ASE) in response to pulsed excitation~\cite{ahn2023electrically}. This is the most significant step towards non-epitaxial laser diodes to date.

Over the last decade, solution-processed metal halide perovskites (MHPs) able to withstand high carrier injection rate have emerged~\cite{kim2018hybrid, kim2020optically, zhao2021nanosecond, elkhouly2022intense, elkhouly2024electrically}.  
 These hybrid organic-inorganic and all-inorganic materials are extensively studied for their optical nonlinearities, whilst offering straightforward and affordable implementation in optical  microcavities~\cite{shi2025ten}. Recently, MHPs demonstrated some of the essential prerequisites for laser diodes: optically pumped continuous-wave polariton lasing~\cite{zou2025continuous} and electrically assisted ASE~\cite{elkhouly2024electrically}, however, several roadblocks remain in the engineering of devices operating exclusively under electrical pumping. In the high current density regime, electroluminescence external quantum efficiency (EL EQE) of vertical structures undergoes roll-off due to charge-carrier imbalance and Joule heating~\cite{kim2018hybrid}. Since perovskites possess low thermal conductivity~\cite{haeger2020thermal}, the heat accumulation promotes non-radiative recombination at room temperature and cannot be avoided even in microstructures combined with heat-dissipative substrates~\cite{elkhouly2024electrically}. Electron-hole imbalance is addressed by employing charge-injection, transport, and blocking organic layers only~\cite{jin2018balancing}. 
 
 The `soft' ionic crystal lattice of MHPs facilitates the drift of its mobile species in external electric field and their reactivity at room temperature. This triggers uncontrollable doping of adjacent functional layers~\cite{zhao2017mobile} and corrosion of metal electrodes~\cite{lin2022electrode}. It is worth noting that cooling down a device to prevent overheating and inhibit unwanted chemical reactions causes an increase in the series resistance of organic layers, and hence can further promote carrier imbalance~\cite{kim2020optically}. These mechanisms prevent perovskite LED architectures to undergo ASE, even at current density exceeding several kA cm$^{-2}$. To date, electrical injection was shown to lower the threshold of optically induced ASE, with  electrical pulses reaching current densities above 3 kA cm$^{-2}$~\cite{elkhouly2024electrically}. These results were reinforced by the observation of electrically assisted optically pumped lasing in a perovskite film under short-, high-current pulses at 230 K~\cite{grede2024electrically}. Recently, Zou et al.~\cite{zou2025electrically} demonstrated a dual-cavity 
device in which a microcavity perovskite LED delivers concentrated optical power into a single-crystal perovskite microcavity to obtain lasing. Such an integrated design is inspired by a well-established technological concept~\cite{yang2008hybrid} that was previously employed for the fabrication of efficient dual-cavity organic LED/organic microcavity laser device~\cite{yoshida2023electrically}. 

A step-change to the required carrier density for achieving lasing comes from driving the electronic transitions into the strong coupling regime, wherein exciton-polaritons, thereafter polaritons, can form in the presence of a high-Q optical cavity. Owing to their bosonic nature, polaritons undergo Bose-Einstein condensation creating a macroscopic coherent state that results in coherent emission from the structure, polariton laser, at carrier densities below the Mott transition~\cite{zou2025continuous,tsotsis2012lasing}. The first demonstration of an electrically pumped polariton laser was implemented on epitaxial structures comprising In$_{0.15}$Ga$_{0.85}$As quantum wells separated by GaAs barriers and integrated with GaAs/AlAs Bragg mirrors~\cite{schneider2013electrically}. A clear transition from incoherent polariton emission to polariton lasing was observed in the presence of an external magnetic field of 5 T.

Here, we report on an electrically driven polariton laser based on a solution-grown CsPbBr$_3$ perovskite microplate, integrated within an optical microcavity. 
Efficient injection of charge carriers into the perovskite is achieved through a chemically inert single-walled carbon nanotube (SWCNT) thin film of electrodes. In the presence of an external electric field, the perovskite's mobile ionic species produce a p-i-n rectifying diode structure that we spatially freeze to optimise the balance of charge carrier injection, through a two-stage cooling of the device to 8 K. We demonstrate polariton lasing with a threshold of 60 $\upmu$A, corresponding to a current density of around 2.8 kA cm$^{-2}$, under continuous electrical injection.

\section{Results and Discussion}\label{sec2}

\subsection{Planar microdevice fabrication}
\begin{figure}[t!]
\centering
\includegraphics[width=0.5\linewidth]{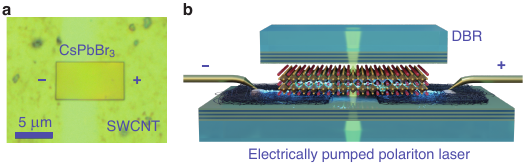}
\caption{$|$\textbf{Architecture of the perovskite polariton laser diode.} \textbf{a}, Optical image of CsPbBr$_3$ microplate on SWCNT non-symmetrical electrodes. \textbf{b}, Schematic illustration of an electrically driven device consisting of a halide perovskite microcrystal intercalated with single-walled carbon nanotube (SWCNT) electrodes and embedded in an optical microcavity.}
\label{Fig1}
\end{figure}

For the synthesis of CsPbBr$_3$ perovskite microcrystals we employ an anti-solvent vapor diffusion method. 2 $\upmu$L droplet of perovskite precursor solution 0.1 M is deposited on SiO$_2$/Al$_2$O$_3$ substrate with island-like morphology and exposed to 2-propanol vapor (1.5 mg cm$^{-3}$) in a sealed Petri dish at 60 $^\circ$C for 5 min. This procedure yields microplates lying on top of alumina islands, see Supplementary Information (SI) section 1~\cite{markina2023perovskite,ermolaev2023giant}. A miniature polydimethylsiloxane (PDMS) lens is used to detach the microcrystals from the substrate~\cite{matchenya2025short}. Electrodes are patterned from a SWCNT thin film on a distributed Bragg reflector (DBR) by laser ablation method~\cite{marunchenko2022single}. Scanning electron microscopy (SEM) image of an inter-electrode space vividly shows SWCNT bundles extending from the electrodes, see SI section 1. Thereafter, we conduct a dry transfer procedure for a perovskite microplate~\cite{matchenya2025short} with lateral dimensions of 5$\times$9 $\upmu$m. The microplate is integrated with the electrodes to create a non-symmetrical device possessing a large-area anode and small-area cathode (Fig.~\ref{Fig1}a). The precise thickness of the microplate equals 435 nm, as measured using atomic force microscopy (AFM), see SI section 1. A top DBR is subsequently applied to the fabricated planar microstructure. We tune the length of the optical microcavity using a custom made metal bracket to apply pressure to the cavity and simultaneously track the position of the bare cavity modes in the spectrum of transmitted and reflected light at a point adjacent to the microplate~\cite{kolker2025room}. The cavity length is adjusted to give approximately 490 nm at cryogenic temperatures, see SI section 2. At the last step, conductive paste is employed to connect SWCNT electrodes with manganin wires enabling electrical measurements under cryogenic conditions. The entire device is schematically illustrated in Fig.~\ref{Fig1}b. Since the both electrodes are made of the same material the resultant structure can be classified as metal-semiconductor-metal (MSM) Schottky diode \cite{sze2021physics, marunchenko2023mixed}. 

\subsection{Carrier injection in a perovskite diode}

\begin{figure}[t!]
\centering
\includegraphics[width=1.0\linewidth]{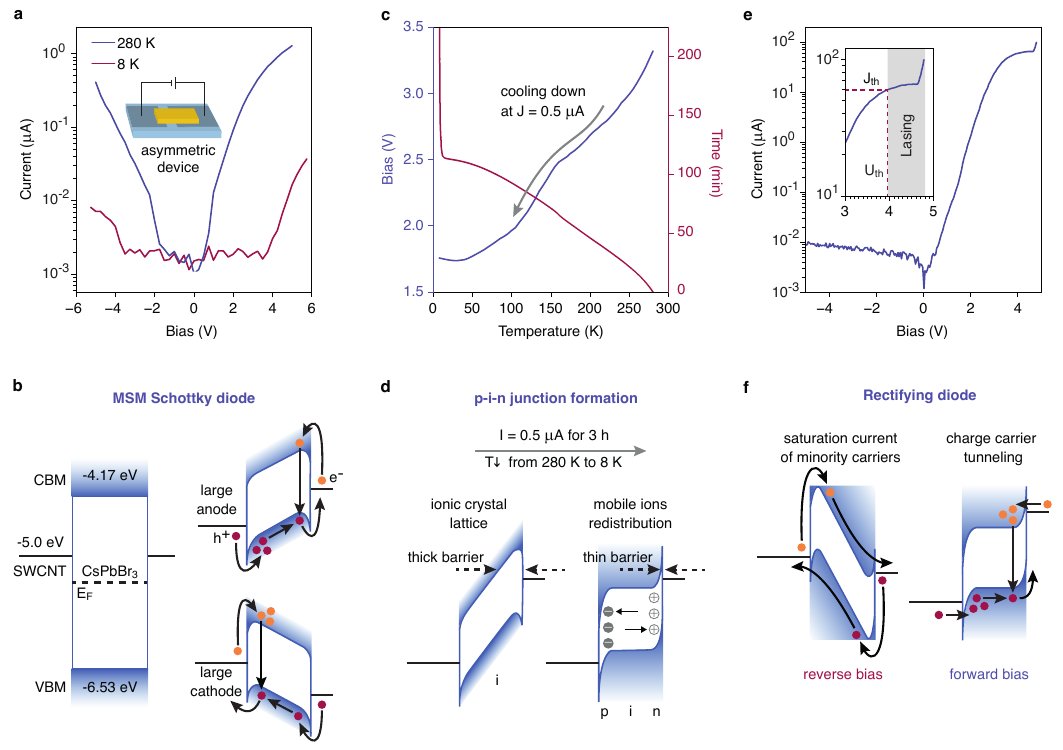}
\caption{$|$\textbf{Charge carrier injection in a perovskite microcrystal.} \textbf{a}, Current-voltage (J-U) curves of metal-semiconductor-metal (MSM) Schottky diode at 280 K and cooled down to 8 K without any electrical stimuli. In the inset the  illustration of the device is presented schematically (top Bragg reflector is omitted for clarity). \textbf{b}, Schematic illustration of carrier injection in MSM-Schottky diode. \textbf{c}, Monotonous decrease in bias of the device during cool down to 8 K at constant current of 0.5 $\upmu$A for 3 h and the corresponding temperature log. \textbf{d}, Formation of a p-i-n junction. \textbf{e}, Asymmetric J-U curve of the device cooled down at constant current and operating at 8 K.  \textbf{f}, Carrier injection energy diagram of a rectifying diode under forward and reverse bias conditions.}
\label{Fig2}
\end{figure}

At room temperature, applied bias naturally induces migration of charged species and transforms the MSM-Schottky diode into an asymmetric p-i-n structure. Mobile ions and their vacancies drift through the `soft' perovskite lattice in the presence of an external electric field toward the electrodes~\cite{eames2015ionic, xiao2015giant}. This redistribution generates p- and n-type regions adjacent to the anode and the cathode, respectively, enhancing the carrier injection rate and thereby promoting electroluminescence (EL) in the perovskite LED without necessitating either charge transport or injection layers~\cite{li2016single}. However, continuous propagation of the p-i-n junction under a constant bias reduces the EL EQE~\cite{shan2017junction}. To circumvent this issue, we test the microcavity at low temperatures. In the absence of any electrical stimuli, we cool the device to 8 K and record  currents not exceeding 40 nA (Fig.~\ref{Fig2}a) due to the significant height of the MSM-Schottky barriers at the SWCNT-CsPbBr$_3$ interfaces, shown in Fig.~\ref{Fig2}b, suggesting that no sufficient redistribution of ions and their vacancies occurs at cryogenic temperatures in the `soft' perovskite lattice. Still, we observe a difference in the current flowing through the structure in the forward and reverse bias configurations and attribute it to the unequal contact areas of perovskite-cathode and perovskite-anode interfaces. These result in different heights for the effective MSM-Schottky barriers, which govern the injection rates~\cite{park2011dark}.

To allow for the formation of a p-i-n junction, whilst preventing its continuous propagation, we adopt a two-stage protocol. At the first stage, we cool the device down to 280 K in order to reduce the ion and vacancy mobility in the `soft' perovskite crystal. At this temperature, we obtain a maximum current of 2 $\upmu$A, that is two orders of magnitude higher than in the MSM-Schottky diode at 8 K (Fig.~\ref{Fig2}a). At the second stage, the device is further cooled down to 8 K at constant current of 0.5 $\upmu$A for 3 h (Fig.~\ref{Fig2}c). This procedure is accompanied with change in bias which monotonously descends in the range of 1.75-3.4 V, see Fig.~\ref{Fig2}c. We attribute this effect to the slow spatial redistribution of the mobile charge species. The redistribution of the ionic species progressively thins the energy barriers at the perovskite-SWCNT interface (Fig.~\ref{Fig2}d) thus enhancing carrier tunneling in the n- and p-type regions~\cite{tung2001recent}. The current-voltage (J-U) characteristics of the in-situ formed p-i-n diode demonstrate a rectifying behavior (Fig.~\ref{Fig2}e). Reverse bias induces a weak saturation current of minority carriers flowing through the p-i-n diode (Fig.~\ref{Fig2}e) due to their thicker energy barriers that inhibite carrier injection (Fig.~\ref{Fig2}f). Conversely, efficient carrier tunneling under forward bias (Fig.~\ref{Fig2}f) yields currents up to four orders of magnitude higher (Fig.~\ref{Fig2}e) than in the MSM-Schottky diode at 8 K (Fig.~\ref{Fig2}a). We note that the formed SWCNT bundles serve as sources of strong electric field at their tips, when bias is applied to the electrodes. The presence of strong electric fields results in the localization of p- and n-type regions. We observe that carrier recombination occurs near the cathode and attribute it to the stronger electric field at the negative side. It is worth noting that this asymmetric electric field naturally appears in perovskite systems governed by one dominant mobile ionic charge and one compensating immobile ion of opposite sign \cite{bertoluzzi2020mobile}. 
            
\begin{figure}[t!]
\centering
\includegraphics[width=1.0\linewidth]{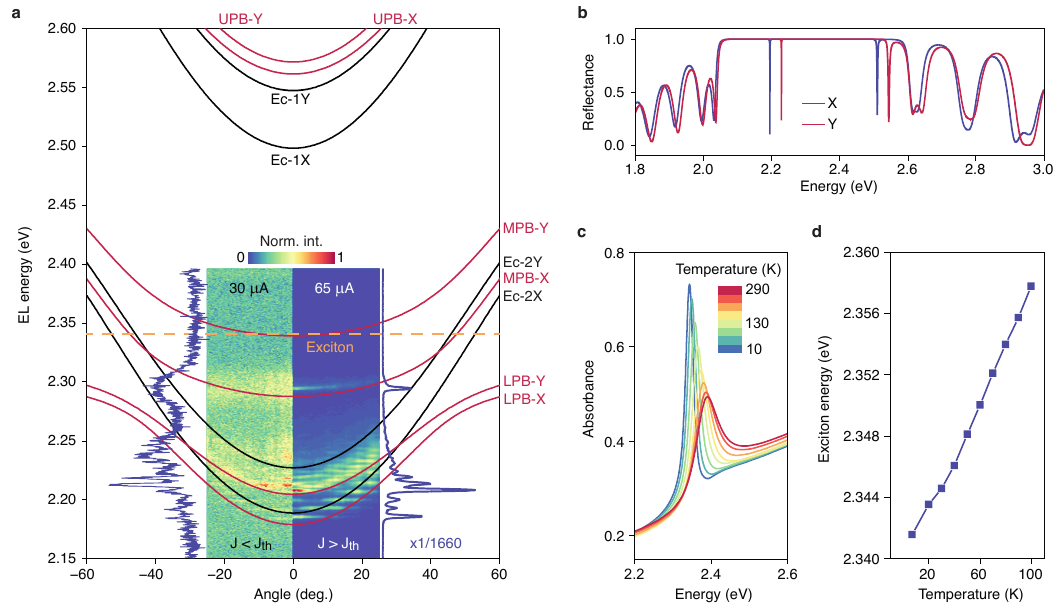}
\caption{$|$\textbf{Strong coupling under direct electrical pumping at 8 K.} \textbf{a}, Measured energy–momentum dispersion and electroluminescence spectra of the microdevice operating at a constant current of 30~$\upmu$A (left panel) and at $65~\upmu$A (right panel). Exciton dispersion is depicted with orange dashed line. Black lines correspond to two pairs of polarisation split cavity modes E$_c$-1X,Y and E$_c$-2X,Y. Red lines correspond to calculated upper (UPB-X,Y), middle (MPB-X,Y), and lower (LPB-X,Y) polariton branches derived from a three coupled oscillators model. Blue lines on the sides represent energy profiles integrated within angle $\pm3^{\circ}$ at 30$~\upmu$A and 65$~\upmu$A respectively. \textbf{b}, Calculated reflectance spectra of 490 nm thick optical microcavity with 435 nm thick CsPbBr$_3$ microplate. X and Y stand for two orthogonal polarisations of the microcavity modes. \textbf{c}, Temperature dependence of the absorbance spectra of a 70 nm thin CsPbBr$_3$ film on a sapphire substrate. \textbf{d}, Exciton energy versus temperature.}\label{fig3}
\end{figure}

\subsection{Polariton condensation under electrical pumping}

We record the energy-momentum dispersion of the electroluminescence under constant current at 8 K by angularly resolving the emission spectrum; for experimental setup see SI section 3. Figure~\ref{fig3}a shows the electroluminescence at 30 $\upmu$A, left color-plot, and 65 $\upmu$A, right color-plot. The corresponding angularly-integrated emission is presented with the blue-spectra to the left and right of the color-plots, respectively. In both dispersions we observe energy discretisation at the lower energy side of the spectrum, whilst a clear narrowing of the spectrum occurs at the emission line at $\approx 2.30$ eV. To understand the nature of the emission spectrum we analyse the energy of the bare modes of the perovskite-containing $2\lambda$ microcavity. Figure~\ref{fig3}b shows the calculated reflection spectrum of the bare cavity modes as derived by the analysis of the reflection and transmission spectrum of the DBRs composing the microcavity. We obtain distinctively different reflectance spectra for X (blue-line) and Y (red-line) polarisations due to the large birefringence of the embedded CsPbBr$_3$ single-crystal~\cite{ermolaev2023giant,kolker2025room}. Within the stop-band we obtain two high-energy (E$_c$-1X,Y) and two low-energy (E$_c$-2X,Y) modes, respectively, see Fig.~\ref{fig3}b, with a pronounced XY-polarisation splitting of $\Delta$E$_{c}$-XY = 33.7 meV. The dispersion of the four bare cavity modes is plotted in Fig.~\ref{fig3}a with black parabolic lines, see SI section 2. Next, we determine the energy of the bare exciton resonance (E$_X$) at 8 K. Figure~\ref{fig3}c shows the temperature dependence of absorption spectrum of a 70 nm bare CsPbBr$_3$ film. We observe that with decreasing temperature, the exciton absorbance spectrum redshifts, narrows and increases in amplitude. We find that the bare exciton energy at 8 K is E$_X$ = 2.341 eV (Fig.~\ref{fig3}d), and annotate it with an orange-dashed line in Fig.~\ref{fig3}a. The dispersion of the bare-cavity and exciton modes are used in a three-coupled-oscillator model for each polarisation, two cavity and one exciton resonance, to obtain the energy-momentum dispersions of the best-fit strongly coupled upper-, middle- and lower-polariton modes (LPB, MPB, UPB), annotated with red-lines in Fig.~\ref{fig3}a, see SI section 4 for calculations.

\begin{figure}[t!]
\centering
\includegraphics[width=1.0\linewidth]{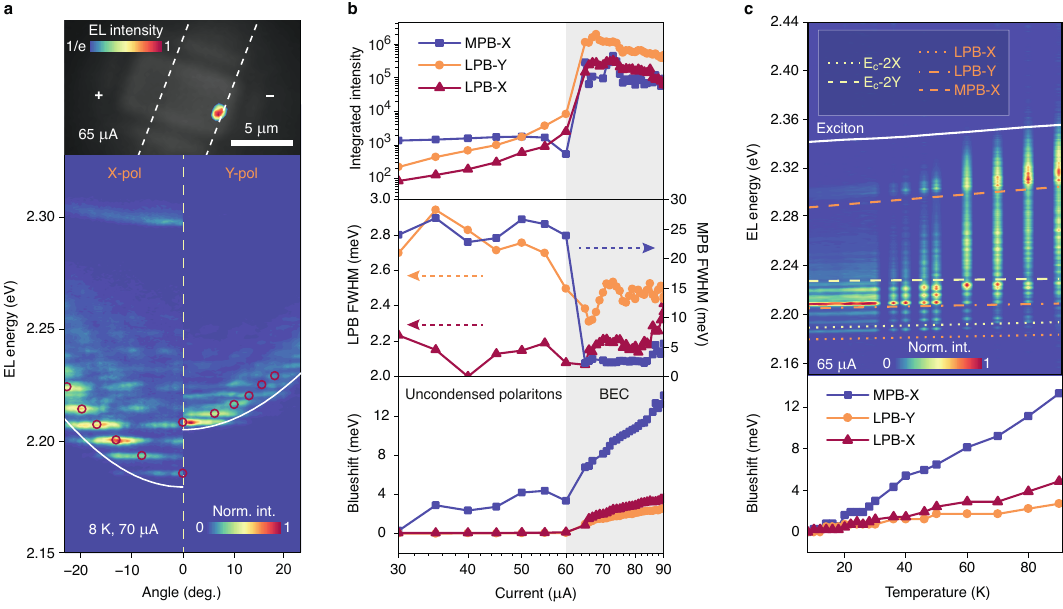}
\caption{$|$\textbf{Polariton lasing and temperature dependence of emission spectrum.} \textbf{a}, Real space image of the electroluminescent center at J = 65 $\upmu$A (top panel). Energy-momentum dispersion for X- and Y-polarisation of the polariton laser at 8 K  at J = 70 $\upmu$A (bottom panel). White solid lines represent the dispersion of lower polariton modes. Discretised energy states of the parabolic potential are marked with red circles. \textbf{b}, Current dependence of the polariton emission integrated within $\pm3^{\circ}$ range for X-polarized MPB and LPB, and Y-polarized LPB. Integrated intensity (top panel), emission line full width at half maximum (FWHM, middle panel), and energy blueshift (bottom panel) are plotted against the injection current. \textbf{c}, Shift of the emission energy integrated within $\pm3^{\circ}$ range versus temperature (top panel). Color and style coded legend depicts the exciton energy and cut-off energy of the cavity modes, and polariton branches. Bottom panel shows the extracted energy blueshift of the three annotated polariton modes versus temperature.}\label{fig4}
\end{figure}

We confirm the presence of the two (X,Y) polarisation modes by performing polarisation-resolved dispersion imaging. Figure~\ref{fig4}a shows the intensity normalised angularly-resolved electroluminescence spectrum for the X- (left color plot) and Y-polarisation (right color plot) at a constant current of 70~$\upmu$A at 8 K. The solid-white lines annotate the calculated lower polariton branches for each polarisation (LPB-X, LPB-Y). We note that for each polarisation, we observe an equidistant energy spectrum (red circles in Fig.~\ref{fig4}a), indicative of a parabolic confining potential for both polarisations ~\cite{Schneider_2017}. From the energy spectrum and the respective effective masses, we estimate that the size of the fundamental states are 0.9 $\upmu$m in diameter for X-, and 1.2 $\upmu$m for Y- polarisations, see SI section 5. We note that the calculated trap sizes are comparable to the dimensions of the real-space electroluminescence pattern $\approx0.79\times1.05
\,\upmu$m, see top panel of Fig.~\ref{fig4}a. 

Following the analysis of the rich energy spectrum observed under electrical pumping, we investigate the linewidth narrowing observed in Figure~\ref{fig3}a between the 30 $\upmu$A and 65 $\upmu$A injection currents by performing a complete injection current dependence of the polarisation-resolved polariton dispersions. Figure~\ref{fig4}b shows the integrated intensity (top panel), the linewidth (middle panel) and the energy blueshift (bottom panel) of the angularly integrated electroluminescence, over $\pm 3^ \circ$, versus the injection current for the three observable polariton modes, LPB-X, LPB-Y and MPB-X. We note here that we do not observe emission from MPB-Y due to strong intracavity losses from its spectral overlap with the absorbance of the bare exciton mode, see relevant energy of bare exciton and MPB-Y in Fig.~\ref{fig3}a. In the top panel of Fig.~\ref{fig4}b we observe a strong increase of emission, exceeding two orders of magnitude, for all three modes, when the current reaches the threshold value J$_{th}$ = 60 $\upmu$A at U$_{th}$ = 3.95 V (Fig.~\ref{Fig2}e). This transition is accompanied by a linewidth narrowing, shown in Fig.~\ref{fig4}b middle panel, and an energy blueshift, shown in Fig.~\ref{fig4}b bottom panel, of all three polariton modes signifying the onset of polariton lasing. 

The nonlinear growth of EL intensity, shown in the top panel of Fig.~\ref{fig4}b, is quickly saturated and then suppressed with further increase of the injection current. We attribute this behavior to self-heating effect which  reduces the EL EQE and triggers impact ionization~\cite{ahmed2018impact} at bias above 4.63 V (Fig.~\ref{Fig2}e). Upon reaching the threshold injection current, a different degree of linewidth narrowing is observed for the three polariton modes, as shown in the middle panel of Fig. \ref{fig4}b. The MPB-X experiences a 10-fold line narrowing from 24 to 2.5 meV, see SI section 6. On the contrary, the FWHM of the LPBs is much narrower even below threshold (2.1~meV for LPB-X and 2.8~meV for LPB-Y) due to the high-Q and large cavity fraction of the LPBs; cavity Hopfield coefficients of $|C|^2=92\%$ for LPB-X and $|C|^2=85\%$ for LPB-Y. Therefore, the linewidth narrowing of LPB-Y is not as dramatic but still narrower than MPB-X (down to 2.3~meV), whereas the narrowing of LPB-X that has the highest cavity fraction is hardly discernible, whilst remaining still narrower than both LPB-Y and MPB-X.

Due to the Wannier-Mott nature of excitons in 3D perovkites, such as the CsPbBr$_3$ microplate here, in the strong coupling regime density-dependent energy shifts are dominated by the repulsive fermionic-exchange nature of polariton-polariton interactions. Below lasing threshold the bottom panel of Fig.~\ref{fig4}b shows a gradual blueshift of the three polariton modes commensurate to their exciton fraction. At threshold current the highly excitonic polaritons of MPB-X ($|X|^{2}=65\%$) demonstrate a bigger energy shift than those of the LPBs. At higher currents, we observe a much steeper increase of the blueshifts in comparison to those below lasing threshold, which are not congruent with the polariton density dependence as inferred from the top panel of Fig.~\ref{fig4}b. 

\par The nature of the energy blueshifts beyond lasing threshold, we attribute to the Joule heating-induced exciton energy-blueshift. To support this assumption, we fix the current at 65 $\upmu$A and perform temperature-dependent, angularly- and spectrally-resolved electroluminescence measurements. The temperature dependence in the range of 8$-$90 K is presented in the top panel color map of Fig.~\ref{fig4}c, wherein we have angularly integrated the emission within an angle range of $\pm3^{\circ}$. On the color plot, we superimpose with lines the temperature dependence of the bare exciton (solid-white), the bare cavity modes (dotted/dashed-white for E$_c$-2X/Y respectively) and the calculated polariton branches at normal incidence (orange lines). The temperature dependence of the energy blueshifts of the three polariton modes, LPB-X, LPB-Y and MPB-X, is shown in the bottom panel of Fig.~\ref{fig4}c. We observe that in the range of 8$-$90 K the energy blueshifts are similar to those observed in the bottom panel of Fig.~\ref{fig4}b, when the current exceeds lasing threshold, reaffirming the presence of Joule heating of the crystal. We note that the carrier density in the luminescence area shown in the top panel of Fig.~\ref{fig4}a remains below the Mott density of n$_M\simeq$10$^{19}$ cm$^{-3}$ in CsPbBr$_3$ at cryogenic temperatures~\cite{ryu2021role}. In particular, the threshold carrier density in the luminescence area is given by n$_{th}$~=~J$_{th}$$\uptau$/qV, where $\uptau=$~40 ps is the lifetime of exciton in bulk CsPbBr$_3$ at cryogenic temperatures~\cite{baranov2020aging}, q stands for the elementary charge, and V is the volume of radiative recombination $\approx$ 0.435$\times$0.79$\times$1.05 $\upmu$m$^{3}$, see SI section 7. Thus, from the threshold current of J$_{th}$ = 60 $\upmu$A to the maximum current of 90 $\upmu$A studied here, the carrier density ranges from n$_{th}$$\sim$4.2$\times$10$^{16}$ cm$^{-3}$ to $\sim$6.2$\times$10$^{16}$ cm$^{-3}$, consistently two orders of magnitude below the Mott density.

\section{Conclusion}\label{sec13}
Three different material classes of solution-processed semiconductors have been contenders for achieving lasing under direct electrical pumping: organic semiconductors, colloidal nanocrystals, and halide perovskites. Here, we demonstrated a perovskite polariton laser diode under continuous direct electrical pumping. The device consists of a metal-semiconductor-metal Schottky diode using single wall carbon nanotube bundles as contact electrodes on a CsPbBr$_3$ microplate embedded in an optical microcavity. We developed a two-stage cryo-cooling protocol to transform the diode into a rectifying p-i-n structure, utilising the 'softness' of the perovskite crystal and transport its ions and their vacancies in the presence of an external electric field. We observed spatially localised polariton electroluminescence and a clear transition to polariton lasing at 60 $\upmu$A. Our discovery introduces a new strategy for engineering solution-processed, electrically-pumped semiconductor lasers, as well as opening the way for real-world applications of polariton lasers.

\bibliography{sn-bibliography}

\makeatletter
\renewcommand\@biblabel[1]{#1.}
\makeatother
\begin{thebibliography}{10}
\expandafter\ifx\csname url\endcsname\relax
  \def\url#1{\burl{#1}}\fi
\expandafter\ifx\csname urlprefix\endcsname\relax\def\urlprefix{URL }\fi
\providecommand{\bibinfo}[2]{#2}
\providecommand{\eprint}[2][]{\url{#2}}
\providecommand{\doi}[1]{\url{https://doi.org/#1}}
\bibcommenthead

\bibitem{shi2025ten}\bibinfo{author}{Shi, Y.} et~al., \bibinfo{title}{Ten years of perovskite lasers}, \emph{\bibinfo{journal}{Advanced Materials}}, \bibinfo{pages}{2413559} (\bibinfo{year}{2025}).
\bibitem{zhu2015lead}\bibinfo{author}{Zhu, H.} et~al., \bibinfo{title}{Lead halide perovskite nanowire lasers with low lasing thresholds and high quality factors}, \emph{\bibinfo{journal}{Nat. Mater.}}, \textbf{\bibinfo{volume}{14}}, \bibinfo{pages}{636--642} (\bibinfo{year}{2015}).
\bibitem{liao2015perovskite}\bibinfo{author}{Liao, Q.} et~al., \bibinfo{title}{Perovskite microdisk microlasers self-assembled from solution}, \emph{\bibinfo{journal}{Advanced materials}}, \textbf{\bibinfo{volume}{27}}, \bibinfo{pages}{3405--3410} (\bibinfo{year}{2015}).
\bibitem{eaton2016lasing}\bibinfo{author}{Eaton, S.~W.} et~al., \bibinfo{title}{Lasing in robust cesium lead halide perovskite nanowires}, \emph{\bibinfo{journal}{Proceedings of the National Academy of Sciences}}, \textbf{\bibinfo{volume}{113}}, \bibinfo{pages}{1993--1998} (\bibinfo{year}{2016}).
\bibitem{tian2023perovskite}\bibinfo{author}{Tian, J.} et~al., \bibinfo{title}{Perovskite quantum dot one-dimensional topological laser}, \emph{\bibinfo{journal}{Nat. Commun.}}, \textbf{\bibinfo{volume}{14}}, \bibinfo{pages}{1433} (\bibinfo{year}{2023}).
\bibitem{kkedziora2024predesigned}\bibinfo{author}{Kedziora, M.} et~al., \bibinfo{title}{Predesigned perovskite crystal waveguides for room-temperature exciton--polariton condensation and edge lasing}, \emph{\bibinfo{journal}{Nat. Mater.}}, \textbf{\bibinfo{volume}{23}}, \bibinfo{pages}{1515--1522} (\bibinfo{year}{2024}).
\bibitem{sun2020lead}\bibinfo{author}{Sun, W.} et~al., \bibinfo{title}{Lead halide perovskite vortex microlasers}, \emph{\bibinfo{journal}{Nat. Commun.}}, \textbf{\bibinfo{volume}{11}}, \bibinfo{pages}{4862} (\bibinfo{year}{2020}).
\bibitem{huang2020ultrafast}\bibinfo{author}{Huang, C.} et~al., \bibinfo{title}{Ultrafast control of vortex microlasers}, \emph{\bibinfo{journal}{Science}}, \textbf{\bibinfo{volume}{367}}, \bibinfo{pages}{1018--1021} (\bibinfo{year}{2020}).
\bibitem{zou2025electrically}\bibinfo{author}{Zou, C.} et~al., \bibinfo{title}{Electrically driven lasing from a dual-cavity perovskite device}, \emph{\bibinfo{journal}{Nature}}, \textbf{\bibinfo{volume}{645}}, \bibinfo{pages}{369--374} (\bibinfo{year}{2025}).
\bibitem{yang2008hybrid}\bibinfo{author}{Yang, Y.}, \bibinfo{author}{Turnbull, G.~A.} \& \bibinfo{author}{Samuel, I. D.~W.}, \bibinfo{title}{Hybrid optoelectronics: A polymer laser pumped by a nitride light-emitting diode}, \emph{\bibinfo{journal}{Applied Physics Letters}}, \textbf{\bibinfo{volume}{92}} (\bibinfo{year}{2008}).
\bibitem{elkhouly2024electrically}\bibinfo{author}{Elkhouly, K.} et~al., \bibinfo{title}{Electrically assisted amplified spontaneous emission in perovskite light-emitting diodes}, \emph{\bibinfo{journal}{Nature Photonics}}, \textbf{\bibinfo{volume}{18}}, \bibinfo{pages}{132--138} (\bibinfo{year}{2024}).
\bibitem{kim2018hybrid}\bibinfo{author}{Kim, H.} et~al., \bibinfo{title}{Hybrid perovskite light emitting diodes under intense electrical excitation}, \emph{\bibinfo{journal}{Nature communications}}, \textbf{\bibinfo{volume}{9}}, \bibinfo{pages}{4893} (\bibinfo{year}{2018}).
\bibitem{kim2020optically}\bibinfo{author}{Kim, H.} et~al., \bibinfo{title}{Optically pumped lasing from hybrid perovskite light-emitting diodes}, \emph{\bibinfo{journal}{Advanced Optical Materials}}, \textbf{\bibinfo{volume}{8}}, \bibinfo{pages}{1901297} (\bibinfo{year}{2020}).
\bibitem{zhao2021nanosecond}\bibinfo{author}{Zhao, L.} et~al., \bibinfo{title}{Nanosecond-pulsed perovskite light-emitting diodes at high current density}, \emph{\bibinfo{journal}{Advanced Materials}}, \textbf{\bibinfo{volume}{33}}, \bibinfo{pages}{2104867} (\bibinfo{year}{2021}).
\bibitem{elkhouly2022intense}\bibinfo{author}{Elkhouly, K.} et~al., \bibinfo{title}{Intense electrical pulsing of perovskite light emitting diodes under cryogenic conditions}, \emph{\bibinfo{journal}{Advanced Optical Materials}}, \textbf{\bibinfo{volume}{10}}, \bibinfo{pages}{2200024} (\bibinfo{year}{2022}).
\bibitem{garcia2017solution}\bibinfo{author}{Garc{\'\i}a~de Arquer, F.~P.}, \bibinfo{author}{Armin, A.}, \bibinfo{author}{Meredith, P.} \& \bibinfo{author}{Sargent, E.~H.}, \bibinfo{title}{Solution-processed semiconductors for next-generation photodetectors}, \emph{\bibinfo{journal}{Nature Reviews Materials}}, \textbf{\bibinfo{volume}{2}}, \bibinfo{pages}{1--17} (\bibinfo{year}{2017}).
\bibitem{saki2021solution}\bibinfo{author}{Saki, Z.}, \bibinfo{author}{Byranvand, M.~M.}, \bibinfo{author}{Taghavinia, N.}, \bibinfo{author}{Kedia, M.} \& \bibinfo{author}{Saliba, M.}, \bibinfo{title}{Solution-processed perovskite thin-films: the journey from lab-to large-scale solar cells}, \emph{\bibinfo{journal}{Energy \& Environmental Science}}, \textbf{\bibinfo{volume}{14}}, \bibinfo{pages}{5690--5722} (\bibinfo{year}{2021}).
\bibitem{deng2022solution}\bibinfo{author}{Deng, Y.} et~al., \bibinfo{title}{Solution-processed green and blue quantum-dot light-emitting diodes with eliminated charge leakage}, \emph{\bibinfo{journal}{Nature Photonics}}, \textbf{\bibinfo{volume}{16}}, \bibinfo{pages}{505--511} (\bibinfo{year}{2022}).
\bibitem{wang2021pursuing}\bibinfo{author}{Wang, K.} \& \bibinfo{author}{Zhao, Y.~S.}, \bibinfo{title}{Pursuing electrically pumped lasing with organic semiconductors}, \emph{\bibinfo{journal}{Chem}}, \textbf{\bibinfo{volume}{7}}, \bibinfo{pages}{3221--3231} (\bibinfo{year}{2021}).
\bibitem{sandanayaka2019indication}\bibinfo{author}{Sandanayaka, A.~S.} et~al., \bibinfo{title}{Indication of current-injection lasing from an organic semiconductor}, \emph{\bibinfo{journal}{Applied Physics Express}}, \textbf{\bibinfo{volume}{12}}, \bibinfo{pages}{061010} (\bibinfo{year}{2019}).
\bibitem{ahn2023electrically}\bibinfo{author}{Ahn, N.} et~al., \bibinfo{title}{Electrically driven amplified spontaneous emission from colloidal quantum dots}, \emph{\bibinfo{journal}{Nature}}, \textbf{\bibinfo{volume}{617}}, \bibinfo{pages}{79--85} (\bibinfo{year}{2023}).
\bibitem{zou2025continuous}\bibinfo{author}{Zou, C.} et~al., \bibinfo{title}{Continuous-wave perovskite polariton lasers}, \emph{\bibinfo{journal}{Sci. Adv.}}, \textbf{\bibinfo{volume}{11}}, \bibinfo{pages}{eadr8826} (\bibinfo{year}{2025}).
\bibitem{haeger2020thermal}\bibinfo{author}{Haeger, T.}, \bibinfo{author}{Heiderhoff, R.} \& \bibinfo{author}{Riedl, T.}, \bibinfo{title}{Thermal properties of metal-halide perovskites}, \emph{\bibinfo{journal}{Journal of Materials Chemistry C}}, \textbf{\bibinfo{volume}{8}}, \bibinfo{pages}{14289--14311} (\bibinfo{year}{2020}).
\bibitem{jin2018balancing}\bibinfo{author}{Jin, X.} et~al., \bibinfo{title}{Balancing the electron and hole transfer for efficient quantum dot light-emitting diodes by employing a versatile organic electron-blocking layer}, \emph{\bibinfo{journal}{ACS applied materials \& interfaces}}, \textbf{\bibinfo{volume}{10}}, \bibinfo{pages}{15803--15811} (\bibinfo{year}{2018}).
\bibitem{zhao2017mobile}\bibinfo{author}{Zhao, Y.} et~al., \bibinfo{title}{Mobile-ion-induced degradation of organic hole-selective layers in perovskite solar cells}, \emph{\bibinfo{journal}{The Journal of Physical Chemistry C}}, \textbf{\bibinfo{volume}{121}}, \bibinfo{pages}{14517--14523} (\bibinfo{year}{2017}).
\bibitem{lin2022electrode}\bibinfo{author}{Lin, C.-H.} et~al., \bibinfo{title}{Electrode engineering in halide perovskite electronics: plenty of room at the interfaces}, \emph{\bibinfo{journal}{Advanced Materials}}, \textbf{\bibinfo{volume}{34}}, \bibinfo{pages}{2108616} (\bibinfo{year}{2022}).
\bibitem{grede2024electrically}\bibinfo{author}{Grede, A.~J.} et~al., \bibinfo{title}{Electrically assisted lasing in metal halide perovskite semiconductors}, \emph{\bibinfo{journal}{ACS Photonics}}, \textbf{\bibinfo{volume}{11}}, \bibinfo{pages}{1851--1856} (\bibinfo{year}{2024}).
\bibitem{yoshida2023electrically}\bibinfo{author}{Yoshida, K.} et~al., \bibinfo{title}{Electrically driven organic laser using integrated oled pumping}, \emph{\bibinfo{journal}{Nature}}, \textbf{\bibinfo{volume}{621}}, \bibinfo{pages}{746--752} (\bibinfo{year}{2023}).
\bibitem{tsotsis2012lasing}\bibinfo{author}{Tsotsis, P.} et~al., \bibinfo{title}{Lasing threshold doubling at the crossover from strong to weak coupling regime in gaas microcavity}, \emph{\bibinfo{journal}{New Journal of physics}}, \textbf{\bibinfo{volume}{14}}, \bibinfo{pages}{023060} (\bibinfo{year}{2012}).
\bibitem{schneider2013electrically}\bibinfo{author}{Schneider, C.} et~al., \bibinfo{title}{An electrically pumped polariton laser}, \emph{\bibinfo{journal}{Nature}}, \textbf{\bibinfo{volume}{497}}, \bibinfo{pages}{348--352} (\bibinfo{year}{2013}).
\bibitem{markina2023perovskite}\bibinfo{author}{Markina, D.~I.} et~al., \bibinfo{title}{Perovskite nanowire laser for hydrogen chloride gas sensing}, \emph{\bibinfo{journal}{ACS nano}}, \textbf{\bibinfo{volume}{17}}, \bibinfo{pages}{1570--1582} (\bibinfo{year}{2023}).
\bibitem{ermolaev2023giant}\bibinfo{author}{Ermolaev, G.} et~al., \bibinfo{title}{Giant and tunable excitonic optical anisotropy in single-crystal halide perovskites}, \emph{\bibinfo{journal}{Nano Letters}}, \textbf{\bibinfo{volume}{23}}, \bibinfo{pages}{2570--2577} (\bibinfo{year}{2023}).
\bibitem{matchenya2025short}\bibinfo{author}{Matchenya, I.} et~al., \bibinfo{title}{Short-term {B}ienenstock-{C}ooper-{M}unro learning in optoelectrically-driven flexible halide perovskite single crystal memristors}, \emph{\bibinfo{journal}{Small Methods}}, \bibinfo{pages}{e00203} (\bibinfo{year}{2025}).
\bibitem{marunchenko2022single}\bibinfo{author}{Marunchenko, A.~A.} et~al., \bibinfo{title}{Single-walled carbon nanotube thin film for flexible and highly responsive perovskite photodetector}, \emph{\bibinfo{journal}{Advanced Functional Materials}}, \textbf{\bibinfo{volume}{32}}, \bibinfo{pages}{2109834} (\bibinfo{year}{2022}).
\bibitem{kolker2025room}\bibinfo{author}{Kolker, M.~D.} et~al., \bibinfo{title}{Room temperature broadband polariton lasing from a cspbbr3 perovskite plate}, \emph{\bibinfo{journal}{Advanced Optical Materials}}, \textbf{\bibinfo{volume}{13}}, \bibinfo{pages}{2402543} (\bibinfo{year}{2025}).
\bibitem{sze2021physics}\bibinfo{author}{Sze, S.~M.}, \bibinfo{author}{Li, Y.} \& \bibinfo{author}{Ng, K.~K.}, \emph{\bibinfo{title}{Physics of semiconductor devices}}  (\bibinfo{publisher}{John wiley \& sons}, \bibinfo{year}{2021}).
\bibitem{marunchenko2023mixed}\bibinfo{author}{Marunchenko, A.} et~al., \bibinfo{title}{Mixed ionic-electronic conduction enables halide-perovskite electroluminescent photodetector}, \emph{\bibinfo{journal}{Laser \& Photonics Reviews}}, \textbf{\bibinfo{volume}{17}}, \bibinfo{pages}{2300141} (\bibinfo{year}{2023}).
\bibitem{eames2015ionic}\bibinfo{author}{Eames, C.} et~al., \bibinfo{title}{Ionic transport in hybrid lead iodide perovskite solar cells}, \emph{\bibinfo{journal}{Nature communications}}, \textbf{\bibinfo{volume}{6}}, \bibinfo{pages}{7497} (\bibinfo{year}{2015}).
\bibitem{xiao2015giant}\bibinfo{author}{Xiao, Z.} et~al., \bibinfo{title}{Giant switchable photovoltaic effect in organometal trihalide perovskite devices}, \emph{\bibinfo{journal}{Nature materials}}, \textbf{\bibinfo{volume}{14}}, \bibinfo{pages}{193--198} (\bibinfo{year}{2015}).
\bibitem{li2016single}\bibinfo{author}{Li, J.} et~al., \bibinfo{title}{Single-layer halide perovskite light-emitting diodes with sub-band gap turn-on voltage and high brightness}, \emph{\bibinfo{journal}{The journal of physical chemistry letters}}, \textbf{\bibinfo{volume}{7}}, \bibinfo{pages}{4059--4066} (\bibinfo{year}{2016}).
\bibitem{shan2017junction}\bibinfo{author}{Shan, X.} et~al., \bibinfo{title}{Junction propagation in organometal halide perovskite--polymer composite thin films}, \emph{\bibinfo{journal}{The journal of physical chemistry letters}}, \textbf{\bibinfo{volume}{8}}, \bibinfo{pages}{2412--2419} (\bibinfo{year}{2017}).
\bibitem{park2011dark}\bibinfo{author}{Park, J.-H.} \& \bibinfo{author}{Yu, H.-Y.}, \bibinfo{title}{Dark current suppression in an erbium--germanium--erbium photodetector with an asymmetric electrode area}, \emph{\bibinfo{journal}{Optics letters}}, \textbf{\bibinfo{volume}{36}}, \bibinfo{pages}{1182--1184} (\bibinfo{year}{2011}).
\bibitem{tung2001recent}\bibinfo{author}{Tung, R.~T.}, \bibinfo{title}{Recent advances in schottky barrier concepts}, \emph{\bibinfo{journal}{Materials Science and Engineering: R: Reports}}, \textbf{\bibinfo{volume}{35}}, \bibinfo{pages}{1--138} (\bibinfo{year}{2001}).
\bibitem{bertoluzzi2020mobile}\bibinfo{author}{Bertoluzzi, L.} et~al., \bibinfo{title}{Mobile ion concentration measurement and open-access band diagram simulation platform for halide perovskite solar cells}, \emph{\bibinfo{journal}{Joule}}, \textbf{\bibinfo{volume}{4}}, \bibinfo{pages}{109--127} (\bibinfo{year}{2020}).
\bibitem{Schneider_2017}\bibinfo{author}{Schneider, C.} et~al., \bibinfo{title}{Exciton-polariton trapping and potential landscape engineering}, \emph{\bibinfo{journal}{Reports on Progress in Physics}}, \textbf{\bibinfo{volume}{80}}, \bibinfo{pages}{016503} (\bibinfo{year}{2016}).
\bibitem{ahmed2018impact}\bibinfo{author}{Ahmed, F.} et~al., \bibinfo{title}{Impact ionization by hot carriers in a black phosphorus field effect transistor}, \emph{\bibinfo{journal}{Nature communications}}, \textbf{\bibinfo{volume}{9}}, \bibinfo{pages}{3414} (\bibinfo{year}{2018}).
\bibitem{ryu2021role}\bibinfo{author}{Ryu, H.} et~al., \bibinfo{title}{Role of the a-site cation in low-temperature optical behaviors of apbbr3 (a= cs, ch3nh3)}, \emph{\bibinfo{journal}{Journal of the American Chemical Society}}, \textbf{\bibinfo{volume}{143}}, \bibinfo{pages}{2340--2347} (\bibinfo{year}{2021}).
\bibitem{baranov2020aging}\bibinfo{author}{Baranov, D.} et~al., \bibinfo{title}{Aging of self-assembled lead halide perovskite nanocrystal superlattices: effects on photoluminescence and energy transfer}, \emph{\bibinfo{journal}{ACS Nano}}, \textbf{\bibinfo{volume}{15}}, \bibinfo{pages}{650--664} (\bibinfo{year}{2020}).
\end{thebibliography}

\end{document}


\title[Article Title]{~~~~~~~~~~Supplementary information
\\
\\
Non-epitaxial perovskite polariton laser diode operating under direct current}


\author*[1]{\fnm{Anatoly P.} \sur{Pushkarev}}\email{anatoly.pushkarev@gmail.com}

\author[1,2]{\fnm{Daria} \sur{Khmelevskaia}}
\equalcont{These authors contributed equally to this work.}

\author[1]{\fnm{Ivan A.} \sur{Matchenya}}
\equalcont{These authors contributed equally to this work.}

\author[1]{\fnm{Stepan A.} \sur{Baryshev}}
\equalcont{These authors contributed equally to this work.}

\author[1]{\fnm{Denis A.} \sur{Sannikov}}

\author[2]{\fnm{Alexey A.} \sur{Ekgardt}}

\author[3]{\fnm{Eduard I.} \sur{Moiseev}}

\author[3]{\fnm{Natalia V.} \sur{Kryzhanovskaya}}

\author[3]{\fnm{Alexey E.} \sur{Zhukov}}

\author[4]{\fnm{Dmitry V.} \sur{Krasnikov}}

\author[5,2]{\fnm{Alexandr A.} \sur{Marunchenko}}

\author[2]{\fnm{Alexey V.} \sur{Yulin}}

\author[4]{\fnm{Albert G.} \sur{Nasibulin}}

\author*[1]{\fnm{Pavlos G.} \sur{Lagoudakis}}\email{p\_lagoudakis@mail.ntua.gr}

\affil[1]{\orgdiv{Hybrid Photonics Laboratory},  
\orgaddress{\street{Bolshoy Boulevard 30, bldg. 1}, \city{Moscow}, \postcode{121205}, \country{Russia}}}

\affil[2]{\orgdiv{School of Physics and Engineering}, \orgname{ITMO University}, \orgaddress{\street{Kronverksky Pr. 49, bldg. A}, \city{St. Petersburg}, \postcode{197101}, \country{Russia}}}

\affil[3]{\orgdiv{International Laboratory of Quantum Optoelectronics}, \orgname{HSE University}, \orgaddress{\street{Soyuza Pechatnikov str. 16}, \city{St.
Petersburg}, \postcode{190008}, \country{Russia}}}

\affil[4]{\orgdiv{Laboratory of Nanomaterials},  
\orgaddress{\street{Bolshoy Boulevard 30, bldg. 1}, \city{Moscow}, \postcode{121205}, \country{Russia}}}

\affil[5]{\orgdiv{Division of Chemical Physics and NanoLund}, \orgname{Lund University}, \orgaddress{\street{PO Box 124}, \city{Lund}, \postcode{22100}, \country{Sweden}}}

\maketitle
\newpage

\tableofcontents

\newpage

\section{Materials, synthesis, and morphology characterization}

Lead (II) bromide (PbBr$_2$, Alfa Aesar, 99.998~$\%$), cesium(I) bromide (CsBr, Sigma-Aldrich, 99.999$\%$), anhydrous dimethyl sulfoxide (DMSO, Sigma-Aldrich, 99.9$\%$), isopropyl alcohol (IPA, technical grade, 95$\%$, Vecton), high-purity aluminum pellets (Ted Pella, Inc, 99.999$\%$), oxalic acid (Vekton, 99.8$\%$) were used as received.

Single-crystal CsPbBr$_3$ microplates were synthesized by using a protocol that is similar to the earlier reported one~\cite{markina2023perovskite}. PbBr$_2$ and CsBr were mixed in a 0.33:0.30 mmol ratio and dissolved in 3 mL of anhydrous DMSO in a N$_2$-filled glovebox with both oxygen and moisture levels not exceeding 1 ppm. 0.03 mmol excess of PbBr$_2$ was employed to prevent the crystallization of Cs$_4$PbBr$_6$ non-perovskite phase. 2 $\upmu$L of the obtained clear solution was drop-casted on a nanostructured Al$_2$O$_3$ substrate at ambient conditions. Thereafter, the substrate with the droplet was sealed in a hot glass Petri dish (100~mL) containing 200~$\upmu$L of IPA and kept at 60~$^\circ$C for 5~min until the droplet was completely dry.

\begin{figure}[h!]
    \centerfirst
    \includegraphics[scale = 1.2]{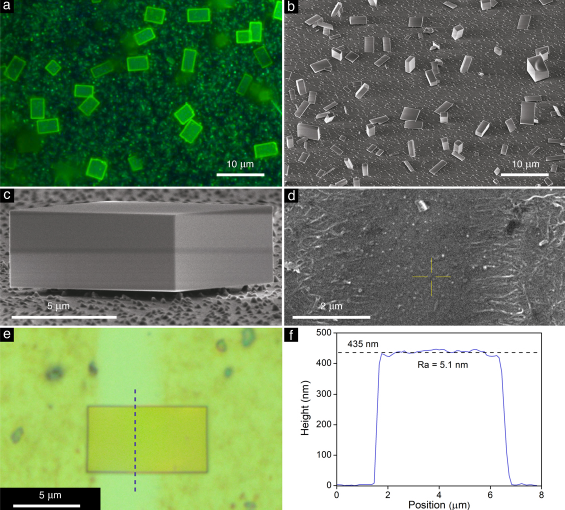}
    \caption{$|$\textbf{Device constituents.} \textbf{a}, Fluorescent image of as-grown CsPbBr$_3$ microplates on nanostructured Al$_2$O$_3$ substrate. \textbf{b}, Large-scale SEM image of microplates on the substrate. \textbf{c},  Tilted-view SEM image of a microplate lying on top of alumina islands. A gap between the bottom face of the crystal and a layer of compact Al$_2$O$_3$ is clearly visualized. \textbf{d}, SEM image of an interelectrode area illustrating residual thin SWCNT bundles at edges of electrodes fabricated by laser cutting of a SWCNT film. \textbf{e}, Optical image of the planar microstructure examined for electrically driven lasing. \textbf{f}, AFM profile of the planar microstructure measured along the dashed line in (\textbf{e}).}
    \label{fig:device_constituents}
\end{figure}

Our technological achievement allowing us to synthesize high-quality perovskite microcrystals and further transfer them to arbitrary substrates for assembling photonic and optoelectronic devices is employment of a substrate with a thin and amorphous Al$_2$O$_3$ layer possessing island-like morphology. This layer was prepared by physical vapor deposition (PVD) of an aluminum film and its further anodizing in oxalic acid solution according the previously published protocol~\cite{markina2023perovskite}. Droplet synthesis of CsPbBr$_3$ perovskite conducted on this nanostructured substrate gave highly luminescent microcrystals (Fig. \ref{fig:device_constituents}a). SEM images of the microcrystals revealed the gap between the alumina compact layer and bottom face of microcrystals (Fig. \ref{fig:device_constituents}b,c). Therefore, the microcrystals were easily detached from the substrate by using a polydimethylsiloxane (PDMS) adhesive lens~\cite{matchenya2025short} and transferred to SWCNT electrodes on DBR.

Laser cutting of SWCNT thing film described in detail in works ~\cite{marunchenko2022single,marunchenko2023mixed,matchenya2025short} produced two separate electrodes (approximate interelectrode distance is 5 $\upmu$m) with the bundles nicely visualized in SEM image (Fig. \ref{fig:device_constituents}d). Before the transfer of perovskite to the electrodes was carried out, 30 V bias was applied to electrodes to burn scarce short circuiting bundles.

Before the assembling of a DBR microcavity, AFM profile of a planar perovskite structure was measured. The mean height of the examined perovskite microcrystal along the dashed line in Fig. \ref{fig:device_constituents}e equals to 435 nm and the mean roughness value is 5.1 nm (Fig. \ref{fig:device_constituents}f)       

\newpage

\section{Engineering of a perovskite-containing DBR microcavity}

\begin{figure}[h!]
    \centering
    \includegraphics[scale = 0.8]{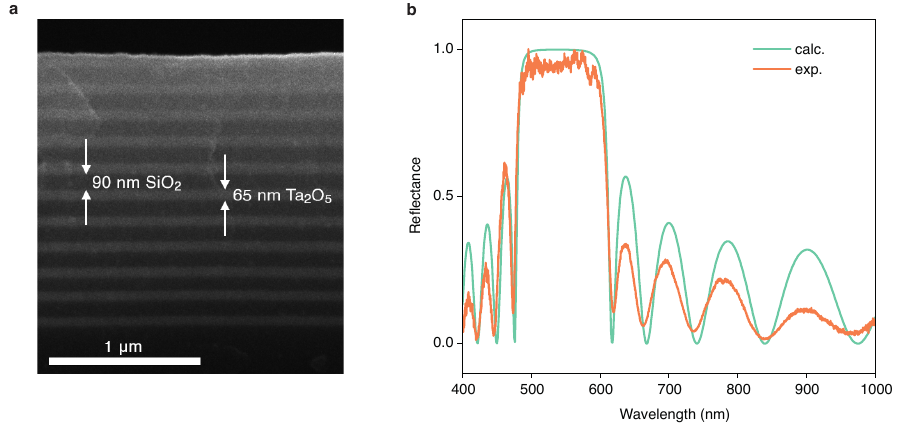}
    \caption{$|$\textbf{Characterization of DBRs.} \textbf{a}, A cross-sectional SEM image of the DBR. \textbf{b} Experimentally measured (orange line) and calculated (green line) reflectance spectra of the DBR.}
    \label{fig:dbr}
\end{figure}

In our experiments, a high-Q microcavity consisted of bottom and top distributed Bragg reflectors (DBRs). The DBRs contain 10.5 pairs of $\approx$~90 nm SiO$_2$ and $\approx$~65 nm Ta$_2$O$_5$ $\lambda$/4 layers (Fig. \ref{fig:dbr}a). The experimental reflectance spectrum from a single DBR is shown in Fig. \ref{fig:dbr}b. The plot demonstrates a stopband in the 485$-$615~nm range.

Numerical calculations of the reflectance and transmission spectra of the single DBR and DBR microcavity were performed by using finite element method in COMSOL Multiphysics software package. A 2D model with periodic boundary conditions (Floquet periodicity) and two periodic input/output ports was employed for the calculations. The material constants of SiO$_2$ and Ta$_2$O$_5$ were taken from \cite{radhakrishnan1951further} and \cite{rodriguez2016self}, respectively. An air (n = 1) was set as a surrounding medium. A plane wave was used as an excitation source. The amplitude reflection and transmission coefficients ($R$ and $T$) were evaluated for different excitation wavelengths from $S$-parameters as $R=|S_{11}|^2$ and $T=|S_{21}|^2$, where $S_{11}$ and $S_{21}$ are the elements of scattering matrix ($S$-matrix).
The calculated reflectance spectrum is in a good agreement with the experimental one (Fig. \ref{fig:dbr}b).

\begin{figure}
    \centering
    \includegraphics[scale = 0.73]{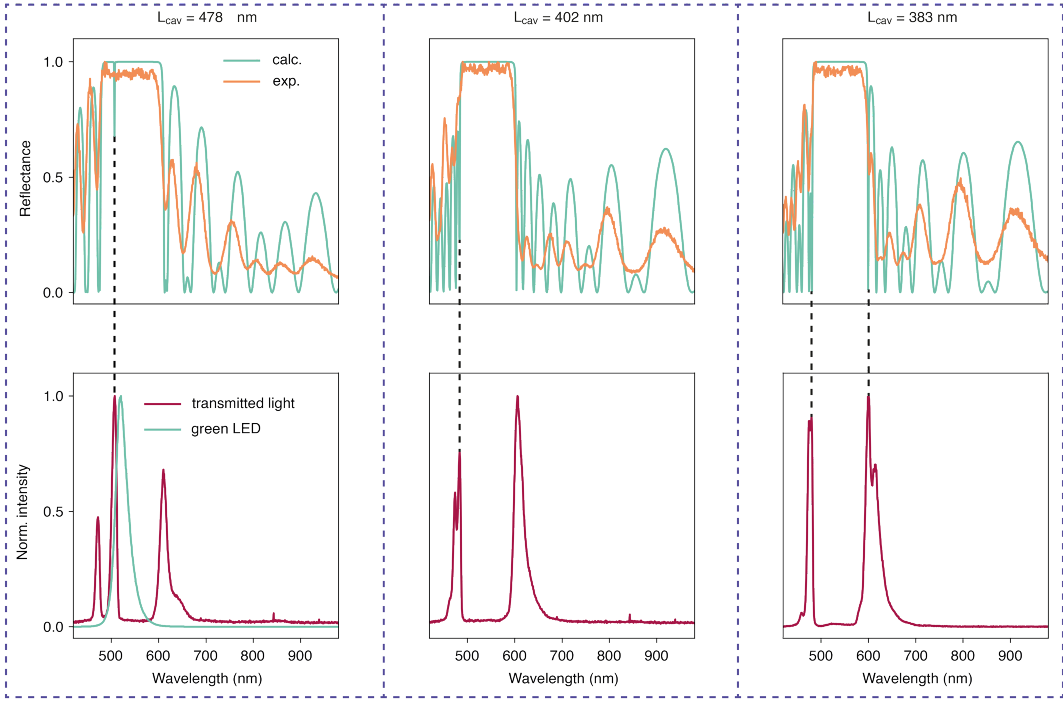}
    \caption{$|$\textbf{Fine tuning of DBR microcavity.}  \textbf{Top panel}, Calculated (green line) and measured (orange line) reflectance spectra for bare cavities of different length. \textbf{Bottom panel}, Experimental spectra of transmitted light (red line) upon broadband green LED illumination (green line). Black dashed lines indicate bare cavity modes matching calculated ones.}
    \label{fig:difcav}
\end{figure}

Fine tuning of the perovskite-containing DBR microcavity length was performed by fitting the calculated reflectance spectra to the experimental one for the bare cavity at the point close to perovskite microcrystal. For this purpose, an air gap between two DBRs varied in numerical simulation. A top panel in Fig. \ref{fig:difcav} illustrates the coherence of the experimental data and calculations for microcavities of different length (L$_{cav}$). To double check the correctness of our approach we measured spectra of transmitted light as well. A bottom panel in Fig. \ref{fig:difcav} shows sharp peaks assigned to bare cavity modes in the corresponding calculated reflectance spectra (marked with dashed black lines). The rest of peaks do not fall within the stopband and stem from broadband green LED illumination.  

\begin{figure}
    \centering
    \includegraphics[scale = 0.8]{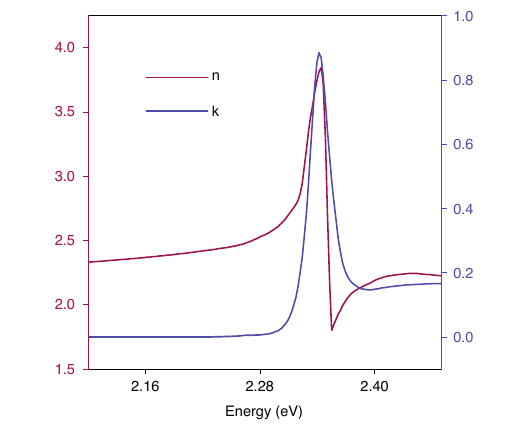}
    \caption{$|$\textbf{Optical constants.} n($\omega$) and k($\omega$) plots at 8 K taken from~\cite{masharin2024polariton}.}
    \label{fig:mat_const}
\end{figure}

By using the aforementioned approach we revealed that in our microdevice kept at 8 K the perovskite microplate  with dimensions of 5$\times$9$\times$0.435 $\upmu$m was confined in a microcavity with L$_{cav}$ = 490 nm. Thereafter, optical properties of the perovskite-containing DBR microcavity were simulated in COMSOL Multiphysics. In numerical simulation, for simplicity, we considered a perovskite medium to be an infinite layer of 435 nm thickness with refractive index n = 2.41 (Fig. \ref{fig:mat_const}) and employed the periodic boundary conditions.

Calculated reflectance spectra for the perovskite-containing DBR microcavity are presented in Fig. 3a in the main text. Y-polarized excitation gives the reflectance spectrum containing two pronounced modes within the stopband having cut-off (at angle 0 deg.) energies E$_{c}$-1Y~$\approx$~2.54~eV and E$_{c}$-2Y~$\approx$~2.23~eV with corresponding Q-factors of $Q_1\approx$~620 (with linewidth of $\delta E_{c}\approx$~4.1~meV) and $Q_2\approx$~6000 ($\delta E_{c}\approx$~0.37~meV)(red line in Fig. 3a). These modes correspond to the different order (N = 5 and N = 4 respectively) Fabry-P$\acute{e}$rot (FP) modes. Since our experimental results showed a significant XY splitting due to anisotropy in a perovskite crystal~\cite{kolker2025room} we introduced a correction to the refractive index $\Delta n$ = 0.07 for numerical modeling of reflectance upon X-polarized excitation. The calculated cut-off energies for the FP modes of the same order are E$_{c}$-1X~$\approx$~2.51~eV with $Q_1\approx$~1590 ($\delta E_{c}\approx$~1.58~meV)  and E$_{c}$-2X~$\approx$~2.19~eV with $Q_2\approx$~4563 ($\delta E_{c}\approx$~0.48~meV) (blue line in Fig. 3a). These values of cut-off energies, linewidths and exciton energy E$_{X}$$\approx$2.341~eV (see Fig. 3c in the main text) were employed as parameters in three coupled oscillators model to fit experimentally observed polariton branches.

\newpage

\section{Experimental setup}

\begin{figure}[h!]
    \centering
    \includegraphics[scale = 0.7]{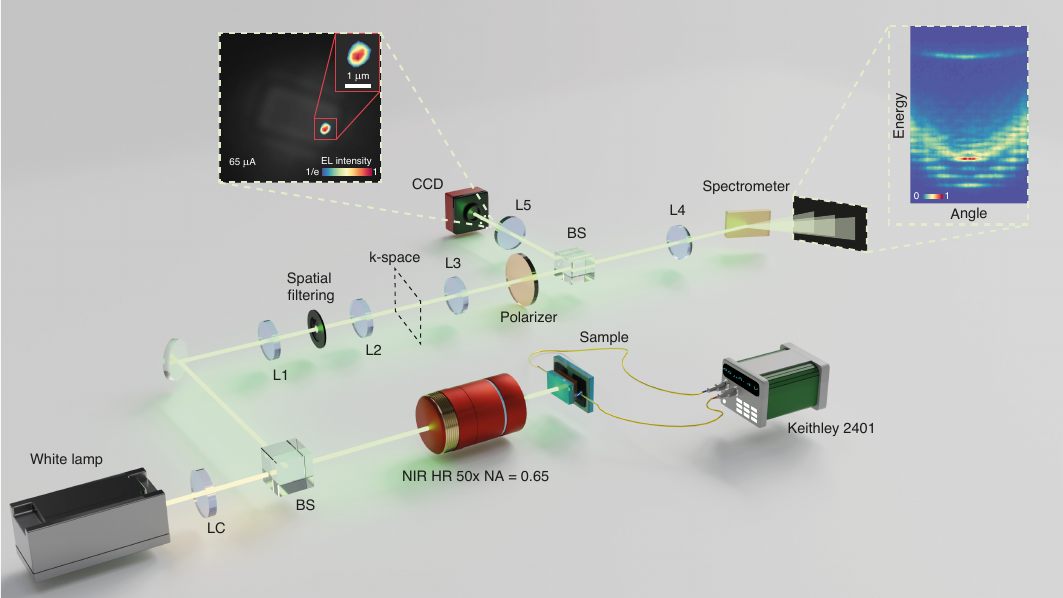}
    \caption{$|$\textbf{Experimental setup.} A setup for the measurement of temperature-dependent electroluminescence (cryostat is omitted for clarity).}
    \label{fig:setup}
\end{figure}

Temperature dependent EL measurements with angular resolution were performed using multifunctional home-built setup depicted in Fig. \ref{fig:setup}. The sample containing perovskite microcavity was installed inside a chamber of closed-cycle helium cryostat (Advanced Research Systems) allowing to tune the temperature in the range of 8$-$300 K. We used Keithley 2401 SourceMeter to electrically excite the sample with constant current. Sample navigation was done upon white light illumination from the halogen lamp Ocean Optics HL-2000. A long-pass filter was used to get rid of the short wavelength light reflected from DBR stopband. The light was focused on the sample using the infinity-corrected objective (Mitutoyo NIR 50x with NA = 0.65). The light reflected from the sample was collected by the same objective, passed through the telescope, consisting of lenses L1 (f=40 cm) and L2 (f=40 cm), and then image of the sample was projected onto the CCD camera with the tube lens. The lens L1 projected the image plane (IP), which allows the spatial filtering of the output signal.

The white light illumination was used to adjust angular-resolved measurements in the reciprocal space (k-space). In the experimental setup, k-space was projected with the lenses L1 and L2 in accordance with 4f scheme \cite{kurvits2015comparative}. The removable lens L3 (f=40~cm) was used in order to image k-space and to transfer it to the CCD camera (Kiralux 1.3 MP NIR-Enhanced CMOS Camera) and to the slit spectrometer coupled to a liquid-nitrogen-cooled imaging CCD camera (Princeton Instruments SP2500+PyLoN). Notably, the presented setup can work in real space without angular distribution if lens L3 is removed.

Pump-dependent electroluminescence spectra with angular distribution was obtained using the same spectrometer with 140 $\upmu$m slit and a 600 grooves/mm grating providing the spectral resolution of around 1.85 meV. Polarization-resolved measurements were performed using a linear polarizer, mounted in circular rotating holder. Exposition time of the spectrometer was set to be 120 s and 1 s below and above polariton lasing threshold, respectively.

\newpage

\section{Three coupled oscillators model}

Our simulations (see Section 2) reveal that for each of polarizations there are two cavity FP modes in the stop-band of a perovskite-containing DBR microcavity. To develop a simple model describing the properties of polariton excitation it is sufficient to take into account the hybridization of the excitons with these two (in each polarization) modes.

The coherent excitons of different spins can be described by the order parameter functions. The excitons of a given spin interact either with clockwise or counter-clockwise polarized light. So the interaction between the linearly polarized photonic mode and the excitons can be written by representing the linearly polarized wave as a combination of two circularly polarized waves. 

We assume that the characteristic times of all processes of interest are much longer compared to the inverse cut-off frequencies of the photonic modes. We also assume that the coupling strengths between the excitons and the photons are also supposed to be much less than the cut-off frequencies. Then the evolution of the photonic field can be described in terms of the slow varying complex amplitudes of the linear photonic modes. 

We consider the case where there are two waves propagating along the long side of the microplate ($x$-axis)  with zero ($y$-polarized wave) and nonzero ($x$-polarized wave) projection of the electric field on the cavity plane to the propagation direction. It is convenient to describe the dynamics in the  Fourier domain ($\omega$-$k$ representation). The equations read

\begin{eqnarray}
\omega A_{x 1}=\left( \delta_{x 1} + D_{x 1}(k) \right) A_{x 1} + \frac{1}{\sqrt{2}}g_{x 1} (\psi_{\uparrow} + \psi_{\downarrow}) \\
\omega A_{x 2}=\left(\delta_{x 2} + D_{x 2}(k) \right) A_{x 2} + \frac{1}{\sqrt{2}}g_{x 2} (\psi_{\uparrow} + \psi_{\downarrow})\\
\omega A_{y 1}=\left( \delta_{y 1} + D_{y 1}(k) \right) A_{y 1} + \frac{1}{i\sqrt{2}}g_{y 1} (\psi_{\uparrow} -\psi_{\downarrow})\\
\omega A_{y 2}=\left(\delta_{y 2} + D_{y 2}(k) \right) A_{y 2} + \frac{1}{i\sqrt{2}}g_{y 2} ( \psi_{\uparrow} - \psi_{\downarrow} )\\
\omega \psi_{\uparrow}=\frac{1}{\sqrt{2}}g_{x 1} A_{x 1}+\frac{1}{\sqrt{2}}g_{x 1} A_{x 2} +\frac{i}{\sqrt{2}}g_{y 1}A_{y 1}+\frac{i}{\sqrt{2}}g_{y 2}A_{y 2} \\
\omega \psi_{\downarrow}=\frac{1}{\sqrt{2}}g_{x 1} A_{x 1}+\frac{1}{\sqrt{2}}g_{x 1} A_{x 2} -\frac{i}{\sqrt{2}}g_{y 1}A_{y 1}-\frac{i}{\sqrt{2}}g_{y 2}A_{y 2} 
\end{eqnarray} 
where $A_{x}$ are the amplitudes of the x-polarized photonic modes, $A_{y}$ are the amplitudes of the $y$-polarized photonic modes, the second index at the amplitudes denotes that it belongs to the lower or the upper photonic mode of the given polarization ($1$ stands for the upper photonic mode and $2$ for the lower one). The indexes $\uparrow$ and $\downarrow$ at the exciton order parameter functions $\psi$ denote the spin of the excitons. 
The detuning of the cut-off frequencies of the photonic modes from the exciton resonance are accounted by the coefficients $\delta$ with the first index denoting the polarization and the second $-$ is the number of the mode. The dispersion of the photonic modes is described by functions $D(k)$ with the indexes having the same meaning as for the amplitudes of the modes. Finally, the interaction of the excitons with the first (upper) photonic mode is accounted by the coefficients $g_{1}$ and with the second (lower) photonic mode $-$ by the coefficients $g_{2}$. 

Introducing the variables  $\psi_{x}=\frac{\psi_{\uparrow}+\psi_{\downarrow}}{\sqrt{2}} $ and $\psi_{y}=\frac{\psi_{\downarrow}- \psi_{\uparrow}}{i\sqrt{2}}$ we can factorize the equations and get two sets of the equations describing the polariton modes of different polarizations. In vector form these equations can be written as
\begin{equation}
\omega \vec Z_{x, \, y} = \hat H_{x, \, y} \vec Z_{x, \, y}  
\end{equation}
where $\vec Z_{x, y}=\left( A_{x, \, y 1}, A_{x,y \, 2}, \psi_{x,y} \right)^T$ and the interactions are described by the operator
\begin{equation}
\hat H_{x,y} = \left( \begin{array}{ccc}
     \delta_{x,y \, 1} + D_{x,y \, 1}(k)  &  0 &g_{x,y \, 1} \\
     0 & \delta_{x,y \, 2} + D_{x,y \, 2}(k) & g_{x,y \, 2} \\
     g_{x,y \, 1} & g_{x,y \, 2} & 0 
\end{array}\right) .
\end{equation}

To link the theory and the experiments we need to find the values of the parameters entering the expression for $\hat H_{x,y}$. Under the conditions discussed above the  photonic modes can be accurately fitted by parabolas, $D_{x,y \, 1,2}(k)=\tilde D_{x,y \, 1,2} k^2$ (each mode has its own dispersion).  Thus, for each of the polarizations we have six parameters ($D_{x,y \, 1}$, $D_{x,y \, 2}$, $\delta_{x,y \, 1}$, $\delta_{x,y \, 2}$,  $g_{x,y \, 1}$, $g_{x,y \, 2}$) to be determined by fitting the experimentally measured dispersion characteristics.

This can be done using the least square method minimizing the discrepancy defined as $\xi = \sum_n ( \omega(k_n)-\omega_{exp \, n} )^2$ where $\omega_{exp \, n}$ are the experimentally measured frequencies corresponding to the wave vectors $k_n$ for the chosen polarization and $\omega(k_n)$ are the frequencies calculated as the eigenvalues of $\hat H_{x}$ or $\hat H_{y}$ for $k_n$ for the corresponding branch of the dispersion characteristic of the polaritons. For $x$-polarization we have reliable experimental data for both polariton branches and use both of them to do the fitting. In the case of $y$-polarized waves we did fit using the experimental data for the lower polariton branch only.  

However it should be taken into account the dependencies of eigenfrequencies $\omega$ of interest on some of the parameters can be weak. In this case it may happen that the fit can give the parameters values significantly deviating from their real values because of the insufficient precision of the experimental measurements and, more importantly,  the imperfection of the theoretical description. To avoid the unphysical results it is worth decreasing the number of fitted parameters. To reduce the parameters numbers we can use the fact that the dispersion coefficient of the photonic mode is linked to its cut-off frequency. Disregarding the material dispersion and the anisotropy we obtain that the frequency of the photonic mode expresses though its in-plane wave vector $k$ and the wavevector perpendicular to the cavity plane $\kappa$ as  $\omega=v\sqrt{\kappa^2+k^2}$ where $v$ is the phase velocity of the waves in the perovskite layer. The condition $k \ll \kappa$ is well satisfied for our experimental conditions and so we can expand the expression for the frequency in Taylor series as $\omega = v\kappa + \frac{v}{2\kappa} k^2$. Then we can estimate the value of $\kappa$ from numerical simulations for the upper and the lower photonic branches. This gives us the relation between the cut-off frequencies of the lower and upper photonic modes and relation of the cut-off frequency to the dispersion of the mode $D=\frac{\omega_0}{2\kappa^2}$ where $\omega_0$ is the cut-off frequency of the mode. This reduces the number of the fitted parameters to just three that can be comfortably fitted by least square method described above. 

\begin{figure}[t]
    \centering
    \includegraphics[scale = 0.7]{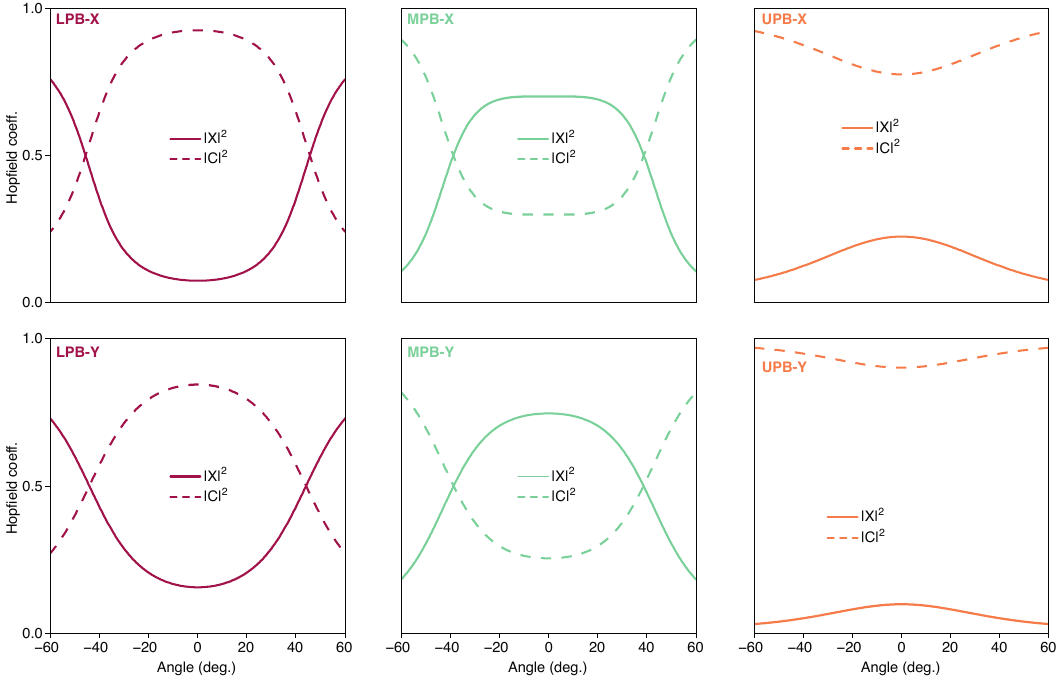}
    \caption{$|$ \textbf{Calculated Hopfield coefficients for LPB, MPB and UPB for both X and Y polarizations.}}
    \label{fig:hopfield}
\end{figure}

The dependencies of the eigenvalues of $\hat H_{x,y}$ on wavevector $k$ give the dispersion characteristic of the lower, middle and upper polariton modes. We provide the best fit results of MPB-X and LPB-X,Y polariton branches shown in Fig. 3d in the main text using aforementioned three coupled oscillators (TCO) model. Notably, the obtained cut-off frequencies for photonic modes are in a good agreement with calculated modes from reflectance/transmittance spectra: $\omega_1\approx$~2.5~eV and $\omega_2\approx$~2.19~eV for X-polarization and $\omega_1\approx$~2.547~eV and $\omega_2\approx$~2.23~eV for Y-polarization, where corresponding dispersion coefficients are D$_x$~=~0.0015 eV$\mu$m$^2$ and D$_y$~=~0.00137 eV$\mu$m$^2$. Obtained polariton modes are in a good agreement with experimental ones (Fig.3d), where the cut-off frequencies of the calculated polariton modes are $\Omega_{MPB-X}\approx$~2.288~eV, $\Omega_{LPB-X}\approx$~2.179~eV and $\Omega_{LPB-Y}\approx$~2.205~eV. The fit reveal following coupling coefficient strength: $g_{x1}=$~117~meV and $g_{x2}=$~34.5~meV, $g_{y1}=$~74~meV and $g_{y2}=$~52.5~meV. The corresponding eigenvectors define the portion of the excitons and the photon belonging to the upper and the lower photonic branches. The Hopfield coefficient defined as the ratio of the exciton to photon fraction is then $K_h=\frac{|Z_3|^2}{|Z_{1}|^2+|Z_2|^2}$ where $Z_1$, $Z_2$ and $Z_3$ are the components of the eigenvector $\vec Z$. The calculated dependence of Hopfield coefficients versus angle is shown in Fig.~\ref{fig:hopfield}.  

Notably, described above three oscillators model can be easily extended to complex eigenvalues when taking into account the linewidths of photonic and excitonic modes. The linewidth of exciton resonance was measured experimentally revealing the value of $\gamma_{exc}\approx$~23~meV at 8 K. The linewidth of the photonic modes was obtained by the Lorentz fitting the calculated transmittance spectra. The Lorentzian fit gives the values of $\gamma_{c,1}\approx$~1.58~meV and $\gamma_{c,2}\approx$~0.48~meV for X-polarization. We employed these values to fit LPB-X and MPB-X and obtained $\gamma_{LPB-X}^{fit}\approx$~2.19~meV and $\gamma_{MPB-X}^{fit}\approx$~17~meV, which is in a respectively good agreement with experimental results ($\gamma_{LPB-X}^{exp}\approx$~2.1~meV and $\gamma_{MPB-X}^{exp}\approx$~24~meV). Similarly, for Y-polarization (with photonic mode linewidth of $\gamma_{c,1}\approx$~4.1~meV and $\gamma_{c,2}\approx$~0.37~meV) we obtained $\gamma_{LPB-Y}^{fit}\approx$~3.9~meV ($\gamma_{LPB-Y}^{exp}\approx$~2.8~meV) and $\gamma_{MPB-Y}^{fit}\approx$~18~meV. According to this, Q-factor for LPB-X was estimated to be around two times higher than one for LPB-Y. Some divergence between experimental and fit result can be caused by optical gain, finite size of the perovskite microplate that were not considered within the model.

\section{Localized polariton states}
To analyze the localized polariton states in a shallow potential it is convenient to work in the polariton basis representing the total field as a combination of polaritons belonging to different branches of the dispersion characteristic. If the depth of the potential is much less compared to the detuning between the branches of the dispersion characteristics then we can assume that the potential causes the mutual scattering of the polaritons belonging to the same dispersion branch only. Then the polariton dynamics can be characterized by a complex slowly varying amplitude $\chi$ (order parameter function) of the considered polariton mode. The equation for the order parameter $\chi$ reads 
\begin{equation}
\partial_t \chi=  i \hat D_{p} \chi + iW(x) \chi,
\end{equation}
where $\hat D_{p}$ is the operator describing the dispersion of the polarions and $W$ is the effective potential acting on the polaritons.
For small wavevectors $k$ the dispersion can be approximated as $\hat D_p = \Omega_p -D_{p0}(\partial_x^2+\partial_y^2)$ where $\Omega_p$ is the cut-off frequency of the polariton mode and $D_{p0}$ is the dispersion coefficient inversly proportional to the polariton effective mass. We assume that the possible anisotropy of the  dispersion can be neglected.  

To explain the equidistant temporal spectrum observed in the experiments we assume that for the low-energy excitations the  effective potential created by the electrodes can be approximated by  a paraboloid with symmetry axis' directed along $x$ and $y$. The directions of the symmetry axis's are defined by the geometry of the electrodes and the applied electric field.    Then the potential can be written in the form  $W=\alpha( x^2  + \mu^2 y^2)$ where a positive parameter $\mu $ accounts for the potential asymmetry.

Under these assumptions the equation for the polariton order parameter $\chi$ reads
\begin{equation}
\partial_t \chi=  i \Omega_p \chi - iD_{p0}(\partial_x^2+\partial_y^2)  \chi + i\alpha( x^2  + \mu^2 y^2) \chi,
\end{equation}
which is a well known equation for two-dimensional quantum oscillator (Schrodinger equation with quadratic potential). 
Looking for the solution in the form $\chi=\chi_{\tilde n}(x) e^{i (\omega_{\tilde n} +\Omega) t}$ we obtain the eigenvalue problem 
\begin{equation}
\omega_{\tilde n} \chi_{\tilde n}=   -D_{p0}(\partial_x^2+\partial_y^2) \chi_{\tilde n} + \alpha( x^2  + \mu^2 y^2) \chi_{\tilde n}. \label{2D_eigvalproblm}
\end{equation}

The eigenfunctions of (\ref{2D_eigvalproblm}) can be found as $\chi_{nm}(x,y)=\varphi_n(x)\phi_m(y)$ where $\varphi$ and $\phi$ are the eigenfunctions of 
\begin{equation}
\omega_n \varphi_n=   -D_{p0} \partial_x^2  \varphi_n + \alpha x^2  \varphi_n, \label{1D_eigvalproblm_1}
\end{equation}
\begin{equation}
\tilde \omega_m \phi_m=   -D_{p0}\partial_y^2 \phi_m + \alpha \mu^2 y^2 \phi_m, \label{1D_eigvalproblm_2}
\end{equation}
corresponding to eigenvalues $\omega_n$ and $\tilde \omega_m$. Here, for sake of convenience, we enumerate the eigenstates $\chi$ by two non-negative indexes $(n, m)$ instead of one index $\tilde n$ ; $\tilde n \rightarrow n,m$. The eigenfrequencies of the state $\chi_{nm}$ is $\omega_{nm}=\omega_n + \tilde \omega_m$. Let us also remark that from (\ref{1D_eigvalproblm_1}),(\ref{1D_eigvalproblm_2}) it is seen that the eigenvales $\tilde \omega_n$ and the eigenfunctions $\phi_n$ are connected with the eigenvales $\omega_n$ and the eigenfunctions $\varphi_n$ by the rescaling $\phi_n(\xi)=\varphi(\sqrt{\mu} \xi)$ and $\tilde \omega_n= \mu \tilde \omega_n$.

From textbooks on quantum mechanics it is known that the eigenfunctions corresponding to 1D quantum oscillators are given by the product of  Gaussian function and Hermitian polynomials $H_n$; 
$\varphi_n=e^{-\frac{x^2}{2 {\cal L}^2}} H_n(\frac{x}{{\cal L}})$ where ${\cal L}=(\frac{D_{p0}}{\alpha})^{1/4}$. The eigenvalues are $\omega_n=\sqrt{D_{p 0} \alpha}(2n+1)$ and  therefore the eigenvalues of the 2D eigenstates are
\begin{equation}
\omega_{nm} = \sqrt{D_{p 0} \alpha}\left( (2n+1) +\mu (2m+1) \right). \label{eig_values}
\end{equation}

Let us notice that for small asymmetry $ |\mu^2-1| \ll 1$ and relatively small indexes $n,m$ the eigenfrequencies of states with  the same $l=n+m$ are close to each other. These states merge into a $l+1$ times degenerate state for $\mu=1$. 

For our experimental conditions we can assume that the asymmetry of the trapping potential is that weak that we cannot spectrally resolve the individual spectral lines and thus we see the spectral maxima at frequencies $\omega_l =\Omega_p +2 \sqrt{D_{p 0} \alpha}(l+1)$. The maximum with $l=0$ contains one individual spectral line, the second maximum $l=1$ contains two individual spectral lines unresolved by the spectrometer and so on.

Analyzing the $l$-th spectral maximum we use the fact that it is formed by $l+1$ eigenstates with similar frequencies. So the field contributing to this spectral peak $\tilde \chi_l$ is the combination of the eigenstates with $n+m=1$:  $\tilde \chi_l=\sum_{m=0}^l \tilde r_{l-m, m} \varphi_{l-m}(x) \phi_{m}(y) e^{i(\Omega+\tilde \omega_m +\omega_{l-m})t}$ where $\tilde r_{n,m}$ are the coefficients reflecting the portion of the $\varphi_n \phi_m$ eigenstate in the total field. 

 In the experiment we measure the spectrum by collecting the radiation coming out along $x$ direction. So what is measured is the spectrum of the field averaged along $y$ direction $\tilde S(k, \omega)=\left| \int \int \int \chi(x,y,t) dy e^{ikx} dx e^{-i\omega t}      dt \right|^2$.  Correspondingly, for the $l$-th spectral peak we obtain  $\tilde S_l(k, \omega)=\left| \int \int \int \chi_l(x,y,t) dy e^{ikx} dx e^{-i\omega t}      dt \right|^2$.  This spectrum can be represented as 
 $\tilde S_l= \sum_{m=0}^l r_{l-m, m} S_{l-m}\delta(\omega-\Omega-\tilde \omega_m - \omega_{l-m}) $ where $\delta$ is Dirac delta-function, $r_{n, m}=|\tilde r_{n, m}|^2 \left|\int \phi_{m}dy\right|^2$ and $S_n=\left| \int \varphi_n e^{ikx} dx \right|^2 $. 
 
 Since we assume that the spectrometer cannot resolve the individual spectral lines then to calculate averaged spatial spectrum $S_{thr \, l}$ of the $l$-th peak theoretically we need to integrate $\tilde S_l$ over the frequencies $S_{thr \, l}=\int \tilde S_l d\omega =\sum_{m=0}^l \frac{1+(-1)^m}{2} r_{l-m, m} S_{l-m}$ Here we took into account that  $r_{n,m}=0$ for odd $m$ because $\phi_m$ in the corresponding integral is then an odd function.

\begin{figure}
    \centering
    \includegraphics[scale = 0.75]{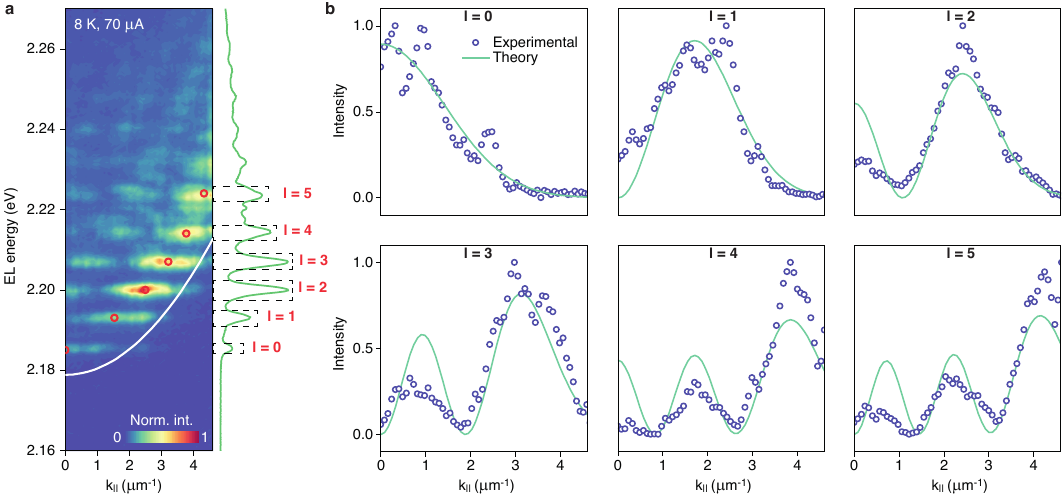}
    \caption{$|$ \textbf{Quantum states of trapped polariton condensate.} \textbf{a}, Calculated quantum states (red circles) of parabolic potential (white solid line) modifying X-polarized LPB. \textbf{b}, Approximation of integrated intensity of each quantum state by the product of a Gaussian function and a Hermitian polynomials.}
        \label{fig:placeholder}
\end{figure}

 From the experiments we know the Fourier representation of the field $S_{exp}(\omega, k)$ shown by color plot in Fig.~\ref{fig:placeholder}a. To find the temporal spectrum we integrate $S_{exp}(\omega, k)$
and obtain the dependency shown in the right part of Fig.~\ref{fig:placeholder}a. From this spectrum we can identify the positions $\omega_l$ of the spectral peaks.   

 To find the experimental spatial spectra for $l$-th peak we average the experimentally measured spectra $S_{exp}(\omega, k)$  over the ranges of frequencies $[\omega_l-\frac{\Delta \omega}{2}; \omega_l+\frac{\Delta \omega}{2} ] $ and obtain $S_{exp \, l}=\int_{\omega_l-\frac{\Delta \omega}{2}}^{\omega_l+\frac{\Delta \omega}{2}}S_{exp}(\omega, k) d\omega$ where $\omega_l$ are the frequencies of spectral maxima and $\Delta \omega$ is the frequency difference between the neighboring spectral maxima. The integration ranges are shown by red dashed lines in  Fig. \ref{fig:placeholder}a. The averaged experimental spatial spectra for first six peaks  $S_{exp \, l}$ are shown in Fig.~\ref{fig:placeholder}b. 
 
 The parameters $D_{p0}$ and $\alpha$ can be found by fitting the experimental spectra $S_{exp \, l}$ by their theoreticl counterparts $S_{thr \, l}=\sum_{m=0}^l \frac{1+(-1)^m}{2} r_{l-m, m} S_{l-m}$.
The spectra $S_{n}$ are the functions of the parameters $\alpha$ and $D_{p 0}$.  As it is shown above the frequency of $l$-th spectral peak is $\omega_l=\Omega_p +2 \sqrt{D_{p 0} \alpha}(l+1)$ and so the 
difference between the neighboring spectral maxima is $2\sqrt{D_{p 0}\alpha}$. This difference is known from the experiments and so we can express $\alpha$ as $\alpha=\frac{\Delta\omega_{exp}^2}{4D_{p0}}$. Thus each of the spectra $S_{nm}(k, D_{p 0})$ becomes a functions of just one unknown parameter $D_{p 0}$. Using the least square method we fit this parameter and the coefficients $r_{nm}$ to get the best agreement between the experimental  $S_{exp \, l}$ and theoretical $S_{thr \, l}$ spatial spectra. These fits are shown in Fig.~\ref{fig:placeholder}b.

The values of $\alpha$ and $D_{p 0}$ found from the first six maxima are presented in the Table \ref{tab:placeholder}.  It is seen that the parameters for the state lowest in energy deviates from the other values significantly. The values of $\alpha$ and $D_{p 0}$ found for the other state have an acceptable deviations from the average value. The average value for the potential parameter $\alpha$ is $<\alpha >=\frac{1}{5}\sum_{n=1}^5 \alpha_n=9.71$ meV/um$^2$ and standard deviation $\Delta \alpha= \sqrt{ \frac{1}{5}\sum_{n=1}^5 (\alpha_n - <\alpha > )^2 }=0.90$ meV/um$^2$. The same values for $D_{p 0}$ are $<D_{p 0}>=1.592$ meV$\cdot$ um$^2$ and $\Delta D_{p 0}=0.136$ meV$\cdot$ um$^2$. 

    \begin{table}
        \centering
        \begin{tabular}{ccccccc}

            index $l$ &      0&  1&  2&  3&  4& 5\\
            \hline
            $\alpha$ in meV/um$^2$  &  $15.61$&   $11.39$&  $9.10$ &  $9.28$ &  $9.91$& $8.88$\\
            \hline
             $D_{p 0}$ in meV$\cdot$ um$^2$ &  $0.982$&  $1.347$&  $1.169$&  $1.654$&  $1.548$& $1.173$\\
             \hline
        \end{tabular}
        \caption{The potential curvature $\alpha$ and the dispersion of the X-polarized polaritons $D_{p 0}$ found from the fit of the experimental averaged spatial spectra $S_{exp \, l}$ by their theoretical counterparts $S_{thr \, l}$. The index $l$ enumerates the temporal spectrum maxima used to calculate the fitted parameters. }
        \label{tab:placeholder}
    \end{table}

A possible explanation on why the values of $\alpha$ and $D_{p 0}$ calculated for the first peak $l=0$ deviate from the values calculated for the other peaks is that the intensity of the first peak is relatively low and this increases the influence of the experimental noise on the presicion of the fit. The second possible reason is that the spatial spectrum corresponding to the first peak has only one maximum situated at $k=0$. The other spatial spectra have more then one maximum and the position of the most pronounced of them can be used to improve the accuracy of the fit.  

\begin{figure}
    \centering
    \includegraphics[scale = 0.75]{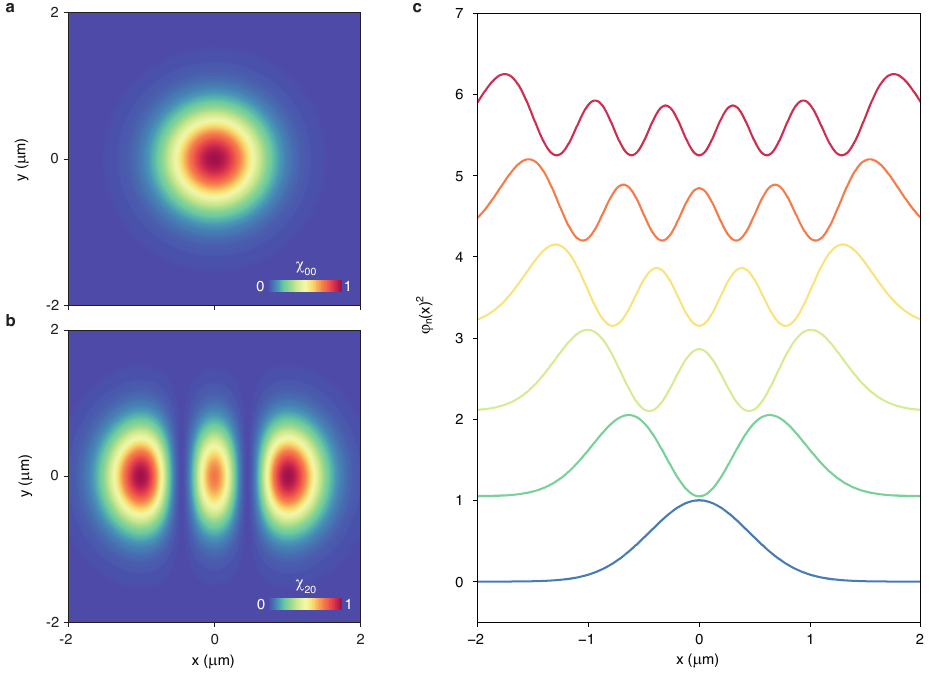}
    \caption{$|$ \textbf{Spatial distribution of the quantum states.} \textbf{a,b}, Calculated field intensity distribution for $\chi_{00}$ and $\chi_{20}$ states. \textbf{c}, Estimated spatial spectra $\phi_n(x)^2$ of the states $\chi_{n0}$, where $n$~=~[0,5].}
    \label{fig:int_states}
\end{figure}

We can also find the expected sized of the eigenstates. For this we calculate the average characteristic scale $<{\cal L}>=\frac{1}{5}\sum_{n=1}^{5} {\cal L}_n=\frac{1}{5}\sum_{n=1}^5 \left( \frac{D_n}
{\alpha_n} \right)^{1/4}$ and using this parameter we can find the field intensity distributions in the states $\chi_{nm}$. For small asymmetry $|\mu-1| \ll 1$ these states can be approximated as $\chi_{nm}=\varphi_n(x)\varphi_m(y)$ 
The fundamental state and the state $m=0$, $n=2$ (which correspond to the brightest peak in  Fig.~\ref{fig:placeholder}a) are shown in Fig.~\ref{fig:int_states}a,b. The dependencies of $\varphi_n(x)^2$ on $x$ are shown in Fig.~\ref{fig:int_states}c for $n=[0, 5]$.

The dispersion coefficient $D_{p 0}=1.592$ meV$\cdot$ um$^2$ fits well to the experimental dependency of the photoluminescence intensity on the frequency and wavevector along $x$-direction (Fig. \ref{fig:placeholder}a) and to the fit obtained from 3-oscillator model, see main text (Fig. 4a). 
The width of the states can be defined as $w=2\sqrt{\int x^2 \varphi_n^2dx /\int \varphi_n^2 dx }$. Using this definition we find the diameter of the fundamental state to be equal to $0.9$ um which is a good estimate for the experimentally observed size of the lasing spot. 

The same fitting can be done for the $y$-polarized states. In this case, as expected, the dispersion of the polaritons is smaller, the average value is about $D_{p0}=1.3$ meV$\cdot$ um$^2$. The parameter characterizing the curvature of the potential is  about $32$ meV/um$^2$. The diameter of the fundamental mode for Y-polarization is close to $1.2$ um.

\newpage

\section{Spectral analysis of MPB-X}

\begin{figure}[h]
    \centering
    \includegraphics[scale = 0.7]{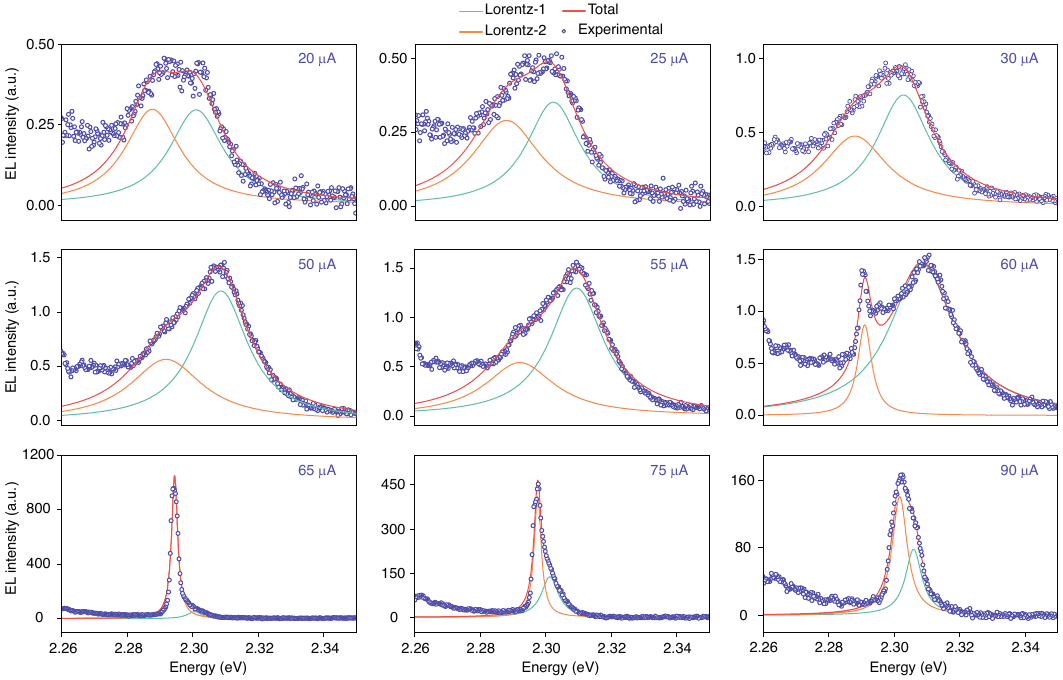}
    \caption{$|$\textbf{EL spectrum evolution of MPB-X with increase in injection current.} Energy profile of MPB-X at different electrical currents integrated within angle $\pm3^{\circ}$. The color coded lines correspond to two Lorentzian fits (green and orange lines) and the sum of the two functions (red line). The experimental data is plotted as blue circles.}
    \label{fig:mpb_fit}
\end{figure}

The energy profile of MPB-X integrated within angle $\pm3^{\circ}$ was measured and fitted. The best approximation was achieved using two Lorentzian functions. FWHM of the emission below the condensation threshold (red line in Fig.~\ref{fig:mpb_fit}) of J$_{th}$ = 60~$\upmu $A was $\operatorname{FWHM_{total}\approx32~meV}$, while the FWHM of the lower energy Lorentzian (orange line in Fig.~\ref{fig:mpb_fit}) function, which exhibits the line narrowing in the main text, was $\operatorname{FWHM_{lor2}\approx24}$ meV.

\newpage

\section{Carrier recombination volume}

\begin{figure}[h!]
    \centering
    \includegraphics[scale = 1]{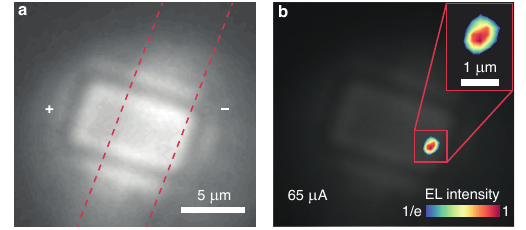}
    \caption{$|$\textbf{Real-space images.} \textbf{a}, Real-space image of the perovskite microcrystal in DBR cavity under white-light illumination. Red dotted lines confine an interelectrode area. \textbf{b}, Normalized polariton emission above threshold (65 $\upmu$A).}
    \label{fig:realspace}
\end{figure}

For the estimation of carrier recombination volume, we employed real-space images of the unbiased (Fig.~\ref{fig:realspace}a) and biased (Fig.~\ref{fig:realspace}b)  microcrystal, respectively. The images show that EL center is localized close to a negative electrode, where SWCNT bundles naturally appear. The lateral dimensions of the emission spot were estimated to be 0.79$\times$1.05 $\upmu$m on the level of 1/e of EL intensity (Fig.~\ref{fig:realspace}). Taking into account the height of the microcrystal h = 0.435 $\upmu$m, the recombination volume equals 0.36$\times$10$^{-12}$ cm$^3$.   

